\documentclass[journal]{IEEEtran}
\IEEEoverridecommandlockouts
\makeatletter
\def\ps@headings{%
\def\@oddhead{\mbox{}\scriptsize\rightmark \hfil \thepage}%
\def\@evenhead{\scriptsize\thepage \hfil \leftmark\mbox{}}%
\def\@oddfoot{}%
\def\@evenfoot{}}
\makeatother
\usepackage{xcolor}
\usepackage{xpatch}

\makeatletter
\usepackage{subfigure}
\usepackage{colortbl}
\usepackage{bm}
\usepackage{float}
\usepackage{graphicx}  % Written by David Carlisle and Sebastian Rahtz
\usepackage{url}       % Written by Donald Arseneau

\usepackage{amsmath}   % From the American Mathematical Society
\usepackage{cite}
\usepackage{colortbl}

\usepackage{amsfonts,amssymb}

\usepackage{stfloats}

\UseRawInputEncoding

\usepackage{cases}
\usepackage{algorithm}
\usepackage{multirow}
\usepackage{algorithmic}
\usepackage{stfloats}%%%% 我加的
\usepackage{epstopdf}%LRY 加
\makeatletter

\newcommand{\Rmnum}[1]{\expandafter\@slowromancap\romannumeral #1@}
\makeatother

\newcommand{\ls}[1]
    {\dimen0=\fontdimen6\the\font
     \lineskip=#1\dimen0
     \advance\lineskip.5\fontdimen5\the\font
     \advance\lineskip-\dimen0
     \lineskiplimit=.9\lineskip
     \baselineskip=\lineskip
     \advance\baselineskip\dimen0
     \normallineskip\lineskip
     \normallineskiplimit\lineskiplimit
     \normalbaselineskip\baselineskip
     \ignorespaces
    }

\newcounter{TempEqCnt}%% 我加的

\begin{document}
\title{RIS-Based Self-Interference Cancellation for Full-Duplex Broadband Transmission}
%\title{Intelligent Reflecting Surface Assisted Full-Duplex Wireless Communication}
%\title{A New SI Cancellation Scheme for Full-Duplex Based on RIS}
%\title{Full-Duplex Self-Interference Mitigation  with Reconfigurable Intelligent Surface }

\vspace{10pt}
\author{\IEEEauthorblockN{Jiayan Wu,~\IEEEmembership{Student Member, IEEE}, Wenchi Cheng,~\IEEEmembership{Senior Member, IEEE},\\ Jianyu Wang,~\IEEEmembership{Student Member, IEEE}, Jingqing Wang,~\IEEEmembership{Member, IEEE}, and Wei Zhang,~\IEEEmembership{Fellow, IEEE}}\\[0.2cm]
\vspace{-10pt}

%\IEEEauthorblockA{State Key Laboratory of Integrated Services Networks, Xidian University, Xi'an, China\\
%E-mail: \{\emph{wccheng@xidian.edu.cn}\}}

\vspace{-25pt}

\thanks{%Part of this work was presented in IEEE International Conference on Communications, 2022\cite{Cao1}.
	This work was supported in part by National Key R\&D Program of China under Grants 2021YFC3002102 and 2023YFC3011502, in part by the Key R\&D Plan of Shaanxi Province under Grant 2022ZDLGY05-09, in part by the National Natural Science Foundation of China under Grant 62201427, and in part by Key Area Research and Development Program of Guangdong Province under Grant 2020B0101110003. {\it{(Corresponding author: Wenchi Cheng.)}}
	
	Jiayan Wu, Wenchi Cheng, Jianyu Wang, and Jingqing Wang are with the State Key Laboratory of Integrated Services Networks, Xidian University, Xi'an, 710071, China 
	(e-mails: jiayanwu@stu.xidian.edu.cn, wccheng@xidian.edu.cn, jywang\_0812@stu.xidian.edu.cn, wangjingqing00@g-\\mail.com).
	
	Wei Zhang is with the School of Electrical Engineering and Telecommunications, University of New South Wales, Sydney,
	NSW 2052, Australia (e-mail: w.zhang@unsw.edu.au).}
}

\maketitle

\begin{abstract}
Full-duplex	(FD) is an attractive technology that can significantly boost the throughput of wireless communications. However, it is limited by the severe self-interference (SI) from the transmitter to the local receiver. %The intelligent reflecting surface (RIS) can intelligently control the wireless environment, which is a plate composed of passive reflect elements. 
%The reconfigurable intelligent surface (RIS) is a plate composed of passive reflect elements, which can intelligently control the wireless environment.
%FD devices enable broadband transmission under frequency-selective fading channels, where power allocation vector are optimized for different subcarriers while only one common set of RIS reflection coefficients is designed to cater to all the subcarriers.
In this paper, we propose a new SI cancellation (SIC) scheme based on reconfigurable intelligent surface (RIS), where small RISs are deployed inside FD devices to enhance SIC capability and system capacity under frequency-selective fading channels. The novel scheme can not only address the challenges associated with SIC but also improve the overall performance.
We first analyze the near-field behavior of the RIS and then formulate an optimization problem to maximize the SIC capability by controlling the reflection coefficients (RCs) of the RIS and allocating the transmit power of the device. The problem is solved with alternate optimization (AO) algorithm in three cases:
ideal case, where both the amplitude and phase of each RIS unit cell can be controlled independently and continuously, continuous phases, where the phase of each RIS unit cell can be controlled independently, while the amplitude is fixed to one, and discrete phases, where the RC of each RIS unit cell can only take discrete values and these discrete values are equally spaced on the unit circle.
%Finally, we utilized alternate optimization (AO) algorithm to jointly optimize the multi-carrier power allocation vector and reflection coefficient (RC) matrix of the RIS in three cases, with the goal of minimizing residual SI.
%For the case of continuous phases, we compare semidefinite relaxation (SDR) and nearest point projection (NPP) methods and show that NPP outperforms SDR.
For the ideal case, the closed-form solution to RC is derived with Karush-Kuhn-Tucker (KKT) conditions. Based on Riemannian conjugate gradient (RCG) algorithm, we optimize the RC for the case of continuous phases and then extend the solution to the case of discrete phases by the nearest point projection (NPP) method.
%We first analyze the near-field behavior of the RIS. Then, we optimize the reflection coefficient (RC) matrix of the RIS to minimize the residual SI in three cases. 
Simulation results are given to validate the performance of our proposed SIC scheme.

%Since the relatively low frequency band has been extensively used, the design paradigm is shifted to the high frequency band such as the promising millimeter wave (mmWave) band in the fifth generation (5G) beyond and the expected terahertz in the sixth generation (6G). Due to the severe electromagnetic degradation for high frequency band, line-of-sight (LOS) transmission for short-range high-speed scenarios has been receiving much attention. In this paper, we introduce the nonplanar electromagnetic waves (NEWs) based wireless communications, which can provide a considerable enhancement for channel capacity. After giving a brief definition of the NEWs, a specific case of the NEWs which is based on the Zadoff-Chu (ZC) phase shift matrix is proposed and analyzed. This kind of NEWs is non-hollow with its orthogonal beams sharing the same radiation pattern except with different rotations along the azimuth. Simulation results are given to validate the capacity enhancement and analyze the robustness performance of NEWs based wireless communications.
\end{abstract}

\vspace{5pt}


\section{Introduction}
%NFC 简介
\IEEEPARstart{F}{ull}-duplex (FD) has gained substantial attention in recent years and is widely considered as an effective technology to enhance the spectrum efficiency. Compared with traditional half-duplex (HD), FD can potentially double the spectrum efficiency, thus inherently  increasing the capacity of wireless networks. However, FD faces the problem of how to deal with the self-interference (SI) \cite{cheng2019full,jimenez2021self,kim2022performance}. 

Reconfigurable intelligent surface (RIS) is an emerging material that consists of a large number of low-cost passive reflecting elements. Moreover, each unit cell of RIS can independently induce the  amplitude and phase variations of the incident signal\cite{tang2021wireless}, thus cooperatively achieving fine 3D reflection beamforming. In the existing literature about RIS-assisted FD communications, the authors of \cite{shen2020beamformig} studied the beamforming optimization for a RIS-assisted FD communication system, where the SI is assumed to follow Gaussian distribution. Also, the authors of \cite{deshpande2022spatially} studied the  spatially correlated RIS-assisted multiuser FD massive multi-input-multi-output (mMIMO) system, where the SI is independent and identically distributed. Although RIS is combined with FD in [5] and [6], how to use RIS to deal with the strong SI was not discussed. The authors of \cite{sharma2020intelligent} proposed a RIS-assisted FD wireless communication system that increases the effective path of the desired signal. However, the self-interference cancellation (SIC) capability is relatively small, because of the high power differential between the desired and interference signals. A RIS-assisted FD system is studied in \cite{dhok2022infinite}, which analyzes the
effects of residual SI, self-reflections and
interference from the RIS. Simulation results of \cite{dhok2022infinite} suggest that RIS can counter the effect of the residual SI and can achieve the same outage probability as that in HD communications if the number of RIS meta-atoms is enough. This suggests the feasibility of utilizing RIS for SIC in FD communications.

On the other hand, with regard to RIS-assisted broadband transmission systems, the authors of \cite{zhang2020capacity} considered the capacity maximization for broadband transmission in a general MIMO orthogonal frequency division multiplexing (OFDM) system under frequency-selective fading channels. In \cite{wei2021sum}, RISs are used in unmanned aerial vehicle (UAV)-based orthogonal frequency division multiple access (OFDMA) communication systems to enhance sum-rate. With a RIS, the composite channel from the UAV to ground users experiences both frequency-selectivity and spatial-selectivity fading. Besides, a RIS-assisted multi-carrier MIMO wireless physical layer security communication system was investigated in \cite{jiang2021joint}, which consists of a
legitimate transmitter, a legitimate receiver, a RIS node, and an eavesdropper. 
It can be observed that while RIS-assisted FD transmission and RIS-assisted broadband transmission have been studied, how to deal with the SI for RIS-assisted FD broadband transmission is not well discussed.
%various applications, such as physical layer secure communications and UAV-based broadband transmission systems, how to cancel SI with RIS is not well discussed.
  %In RIS-assisted FD-OFDM system, RIS brings the same phase shift on different subcarriers. 

%There is limited research on deploying RIS inside devices for FD communications, while
To achieve miniaturization and portability of RIS-assisted FD devices, RISs are desired to be deployed inside the devices. Similar studies have been conducted in the area of visible light communication (VLC). 
In VLC, RIS can be deployed inside devices
to enhance their performance. For example, \cite{aboagye2022ris} depicts an indoor VLC system where RIS assists in directing the received light and concentrating its beam on the photodetector (PD). It functions as a filter that permits light signals of specific wavelengths to pass while impeding unwanted light, resulting in reduced interference. The authors of \cite{aboagye2022design} proposed and demonstrated that deploying liquid crystal (LC)-based RIS as part of the VLC receiver can improve data rate. The purpose
of LC-RIS is providing incident light steering and intensity amplification to improve the received signal strength and the corresponding achievable data rate. The indoor VLC system is composed of a VLC-enabled ceiling light emitting diode array and a user equipped with a nematic LC RIS-based VLC receiver. 
The feasibility of miniaturized RISs with similar size to the RIS in this paper has also been demonstrated in existing works. For examples, the authors of~\cite{yang2022power} proposed and demonstrated the use of a 6-by-8 unit metasurface to protect sensitive devices from large signals, while still allowing a communication channel for small signals. The authors of~\cite{baghel2019linear} fabricated a novel ultrathin linear-to-cross-polarization transmission converter based on a $80 \times 80 ~\rm mm^2$ metasurface (having $10 \times 10$ unit cells).

%These findings provide indirect evidence for the effectiveness of deploying RIS in full-duplex communication systems and underscore the potential of RIS for mitigating self-interference in broadband transmission.

Inspired by the above works related to VLC where RISs can be deployed inside devices to enhance performance, in this paper we consider an FD broadband wireless communication system where small RISs are deployed inside FD devices to cancel SI. With the proposed design for the reflection coefficient (RC) matrix of the RIS, the SI can be significantly canceled, which promises the double data rate as compared with HD. The main contributions of this paper are summarized
as follows:
\begin{itemize}
	\item[1)]
	To the best of our knowledge, this is the first work to study RIS-assisted FD-OFDM communication system where small RISs are deployed inside FD devices to mitigate the strong SI by passive beamforming. The system consists of two FD devices and each device is equipped with a single transmit antenna, a single receive antenna, and a RIS.
	The RIS-based SIC scheme developed in this paper is a new attempt of SIC in the propagation domain. Furthermore, the deployment of small RISs within communication terminals showcases a new application scenario for RIS-assisted wireless communications.
\end{itemize}
\begin{itemize}
	\item[2)]
	%Since small RISs are deployed inside device terminals, the RIS are positioned very close to the transmit antenna and receive antenna, the near-field behavior of the RIS need to analysis and takes into account the near-field behavior of the RIS when analysis the near-field channel of RIS. thus use an exact model to quantify the near-field channel of RIS for broadband transmission.
	Since small RISs are deployed inside devices, it is essential to account for the near-field behavior of the RIS. %To achieve this, an exact model is used to accurately quantify the near-field channel of the RIS, ensuring its effectiveness for broadband transmission in this scenario.
	%The near-field channel model was utilized to accurately model the entire system, ensuring that the simulation results were more realistic.
	RIS-assisted FD-OFDM system is accurately modeled by the near-field channel model, which ensures its effectiveness for broadband transmission in this scenario.
\end{itemize}
\begin{itemize}
	\item[3)]
	To enhance the performance of the new SIC scheme for broadband transmission, we formulate an optimization problem to maximize the SIC capability.
	Given the non-convexity of the optimization problem, an alternate optimization (AO) algorithm is developed to jointly optimize the power allocation vector and RC matrix in three cases. Specifically, for the fixed RC, the closed-form solution to power allocation is derived. For the fixed power allocation vector, quadratic transform, Karush-Kuhn-Tucker (KKT) conditions, Riemannian conjugate gradient (RCG) algorithm, and nearest point projection (NPP) method are adopted to optimize the RC matrix.%, thereby obtaining the optimal RIS phase shift to enhance the SIC capability as much as possible.
	 %Because of the non-convex optimization problem, an alternate optimization (AO) algorithm is designed to jointly optimize the multi-carrier power allocation vector and RC matrix of the RIS in three cases. To be specific, given the fixed RC of the RIS, the close-form solutions of power allocation are derived, given the fixed power the multi-carrier allocation vector, the quadratic transform, semidefinite relaxation (SDR) method, and nearest point projection (NPP) method are adopted again to optimize RC matrix of the RIS to obtain the optimal RIS phase shift as much as possible.
\end{itemize}
\begin{itemize}
	\item[4)]
	Finally, numerical results are presented to validate the performance of the proposed SIC scheme and the effectiveness of the RIS phase shift design. Numerical results demonstrate that the capacity gain of RIS-assisted FD-OFDM system over HD-OFDM system without RISs can be more than 2.25 times.
	%Finally, numerical results are presented to validate the performance of the proposed SIC scheme and the effective of the design of RIS phase shift which can significantly mitigation the strong SI, and then let the capacity gain of the proposed RISs-assisted FD-OFDM communication system over HD-OFDM system without RISs can be more than 2.3 times. 
\end{itemize}

%Next, we consider capacity maximization for broadband transmission in a general orthogonal frequency division multiplexing (OFDM) system under frequency-selective fading channels, where transmit covariance matrices are optimized for different subcarriers while only one common set of RIS reflection coefficients is designed to cater to all the subcarriers. 

\begin{figure}[t]
	\centering
	\vspace{3pt}
	\includegraphics[scale=0.75]{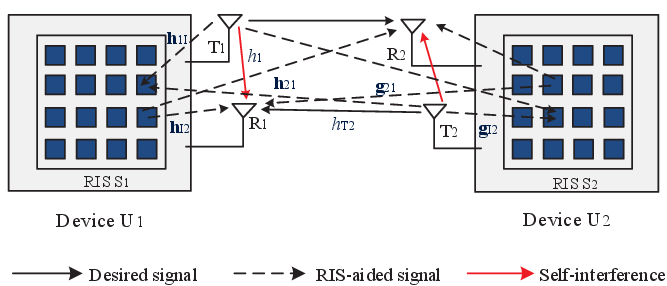}
	\caption{RIS-assisted FD-OFDM communication system.} \label{fig:System_model}
	\vspace{-10pt}
\end{figure} 
\par The rest of the paper is organized as follows. Section~\ref{sec:System_model} gives the system model for the proposed RIS-based SIC scheme and analyzes the near-field behavior of the RIS. In Section~\ref{sec:Residual SI Minimization}, an optimization problem is formulated to maximize the SIC capability. Then we utilize AO algorithm to jointly optimize the power allocation vector and RC matrix in three cases, with the goal of minimizing residual SI.
%Specifically, for the case of continuous phases, we employed SDR method and NPP method to optimize the RC of the RIS, respectively.
%we optimize the reflection coefficient (RC) matrix of the RIS to minimize the residual SI in three cases. 
Simulation results are given in Section~\ref{sec:Simulation} to verify the performance of the proposed SIC scheme. This paper concludes with Section~\ref{sec:Conclusion}.

\textit{Notations}: Boldface straight lowercase and uppercase letters denote vectors and matrices, respectively. %${\rm A}_{\xi,\varsigma}$ denotes the $\xi$th row and $\varsigma$th column element of a matrix $\boldsymbol{\rm A}$. $\boldsymbol{\rm A}_{\xi,:}$ and $\boldsymbol{\rm A}_{:,\varsigma}$ denote the $\xi$th row and $\varsigma$th column of $\boldsymbol{\rm A}$, respectively.
 $\mathbb{C}^{x \times y}$ denotes the space of
$x\times y$ complex matrices, $\mathbb{R}^{x \times y}$ denotes the space of $x\times y$ real matrices, and
$\max(x, y)$ denotes the maximum between two real numbers
$x$ and $y$.
  For a matrix $\rm C$ of arbitrary size, $\boldsymbol{\rm C}^{\rm H}$, $\parallel\boldsymbol{\rm C}\parallel $, and $\boldsymbol{\rm C}^{\ast }$ denote the conjugate transposed matrix, the Frobenius norm, and the conjugate of $\boldsymbol{\rm C}$, respectively. Further, we use
  $\boldsymbol{\rm I}_{N}$ and $\boldsymbol{\rm 0}_{N}$ to denote an identity matrix and all-zero matrix of
  appropriate dimensions, respectively. $\triangleq$ and $\sim$ stands for ``defined as'' and ``distributed
  as'', respectively. %$\boldsymbol{\rm A}^+$ denotes the pseudo-inverse of $\boldsymbol{\rm A}$, and ${\rm \alpha}_\xi$ denotes the $\xi$th element of a vector $\boldsymbol{\rm \alpha}$.
$\bmod (\cdot, \cdot)$ denotes the modulo operation, $\left\lfloor \cdot  \right \rfloor$ denotes round down, and $\odot$ denotes Hadamard product. The distribution of a
circularly symmetric complex Gaussian (CSCG) random variable with mean $\mu $ and variance $\sigma^{2}$ is denoted by $\mathcal{CN}(\mu ,\sigma^{2})$. $\bf c\sim \mathcal{CN}(\mu ,\bf C)$ indicates a complex random vector $\bf c$ 
following the distribution of mean $\mu $ and covariance matrix $\bf C$. $\left|\cdot\right|$ denotes the absolute value operation, $\mathrm{tr}(\cdot)$ denotes the operation of trace, $\tan ^{-1}(\cdot)$ denotes the arctangent function, and $\rm{diag}\left(\mathbf{c}\right)$ denotes a square
diagonal matrix with the elements of $\mathbf{c}$ on the main diagonal. $\arg (\cdot)$ and $\Re(\cdot)$ denote the angle and the real part of a complex number, respectively.

\newcounter{TempEqCnt1} % ??????TempEqCnt
\setcounter{TempEqCnt1}{\value{equation}} % ??????? ??TempEqCnt
\setcounter{equation}{2}
\begin{figure*}[hb]
	\rule[-1pt]{18.1cm}{0.07em}
	\begin{align}
	\xi_{m,\mathbf{t}, \mathbf{p}_{m,n}}=\frac{1}{4 \pi} \sum_{x \in \mathcal{X}_{m,t, n}} \sum_{y \in \mathcal{Y}_{m,t, n}}\left(\frac{{x y}/{d^{2}}}{3\left({y^{2}}/{d^{2}}+1\right) \sqrt{{x^{2}}/{d^{2}}+{y^{2}}/{d^{2}}+1}}+\frac{2}{3} \tan ^{-1}\left(\frac{{x y}/{d^{2}}}{\sqrt{{x^{2}}/{d^{2}}+{y^{2}}/{d^{2}}+1}}\right)\right)
	\label{eq_channel}
	\end{align}
\end{figure*}
\setcounter{equation}{\value{TempEqCnt}}

\section{System Model}\label{sec:System_model}

%we consider an RIS-assisted FD-OFDM communication system under frequency-selective fading channels where two FD devices ($ U_1, U_2$) exchange their information with each other, as illustrated in Fig.~1. 
As shown in Fig. 1, we consider an FD-OFDM communication system where two FD devices (denoted by $U_1$ and $U_2$) exchange information with each other under frequency-selective fading channels.
Both FD devices have a single transmit and receive antenna. We assume that each device is equipped with a RIS which contains $N$ reflection elements to assist FD wireless communications. The RIS assists the FD device through its  $N$ reflection elements which could control the amplitude and phase of the incident signal independently. The receive antenna receives not only the SI signal but also the reflected signal, which is utilized to diminish the SI signal. %in anti-phase of the SI signal.
Thus, the SI can be significantly canceled.

The reconfigurability of RIS lies not only in the amplitude and phase, but also in frequency and polarization shifts~\cite{9994338}. In this paper, the RIS is deployed inside the device, and the positions of the transmit antenna, the receive antenna, and the RIS are fixed. While the RIS channel gain is not affected by the distance between the antennas and the RIS, this paper considers a wideband OFDM system where the RIS channel gain is also affected by the frequency. The RIS channel varies with the frequency, and if the bandwidth exceeds the available bandwidth, it is necessary to reconfigure the wireless environment to ensure the performance gain of the RIS, which reflects the reconfigurability of RIS~\cite{zhang2018space}. 
 
\begin{figure}[t]
	\centering
	\vspace{3pt}
	\includegraphics[scale=0.55]{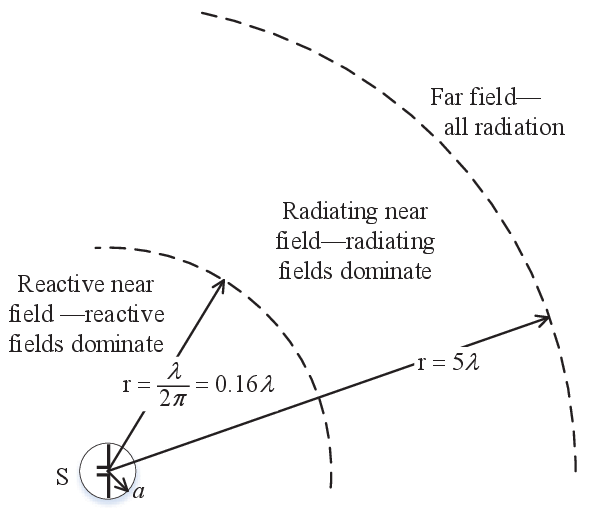}
	\caption{The definition of near and far-field for a small antenna (a sphere with radius $a \ll \lambda$).} \label{fig:definition_nearfield}
	\vspace{-10pt}
\end{figure}
\par We consider the case where the operating frequency is $f=5.8$ GHz\cite{green2022optically} and the wavelength is $\lambda=c/f=0.0517$~m. Also, RISs are considered small and deployed inside FD devices. The distance between the transmitter/receiver of each FD device and each unit cell of its equipped RIS is assumed to be larger than $0.16\lambda$ and less than $5\lambda$. According to the definition of the near-field and far-field  \cite{stutzman2012antenna} for the small-size antenna $S$ shown in Fig. 2, the transmitter/receiver of each FD device is in the radiating near-field of each unit cell of its equipped RIS. The near-field behavior and far-field behavior of the RIS are different. There are three differences between the near-field and the far-field \cite{de2020near}: the distance to each unit cell, the effective unit cell areas, and the loss from polarization mismatch.
Considering the above three differences, in the following the near-field behavior of the RIS is analyzed and utilized to mitigate SI. 
 
\subsection{Channel Model}
Based on previous RIS-assisted FD-OFDM system
model, the channel model is introduced in this subsection. We define the number of OFDM subcarriers as $M$ and the indexes of subcarriers are denoted by $\mathit m\in \mathcal{M} \triangleq\{0,1, \ldots, M-1\}$. Thus the subcarrier spacing $\Delta f=B/M$, where $B$ denotes the transmission bandwidth. We let $\mathit h_{m,\rm 1}$ and $\mathit g_{m,\rm 1}$ denote the SI channel gains of device $U_1$ and device $U_2$ on the $m$th subcarrier, respectively. %Let $\mathit{h}_{m,\rm T2}$, $\mathbf {g}_{m,21} \in \mathbb{C}^{N \times 1}$, and $\mathbf{g}_{m,12} \in \mathbb{C}^{N \times 1}$ denote the frequency-domain channel from transmit antenna  $T_2$ to receive antenna $R_1$, from RIS $S_2$ to receive antenna $R_1$, and from transmit antenna  $T_1$ to RIS $S_2$ on the $m$th subcarrier, respectively.
The frequency-domain channel from transmit antenna  $T_2$ to receive antenna $R_1$, from RIS $S_2$ to receive antenna $R_1$, and from transmit antenna  $T_1$ to RIS $S_2$ on the $m$th subcarrier are denoted by $\mathit{h}_{m,\rm T2}$, $\mathbf {g}_{m,21} \in \mathbb{C}^{N \times 1}$, and $\mathbf{g}_{m,12} \in \mathbb{C}^{N \times 1}$, respectively. 
%Let $\mathit {h}_{m,\rm T1}$, $\mathbf{h}_{m,12} \in \mathbb{C}^{N \times 1}$, and $\mathbf{h}_{m,21} \in \mathbb{C}^{N \times 1}$ denote the frequency-domain channel from transmit antenna $T_1$ to receive antenna $R_2$, from RIS $S_1$ to receive antenna $R_2$, and from transmit antenna $T_2$ to RIS $S_1$ on the $m$th subcarrier, respectively. 
The frequency-domain channel from transmit antenna $T_1$ to receive antenna $R_2$, from RIS $S_1$ to receive antenna $R_2$, and from transmit antenna $T_2$ to RIS $S_1$ on the $m$th subcarrier are denoted by  $\mathit {h}_{m,\rm T1}$, $\mathbf{h}_{m,12} \in \mathbb{C}^{N \times 1}$, and $\mathbf{h}_{m,21} \in \mathbb{C}^{N \times 1}$, respectively. 
The frequency-domain channel between transmit antenna $T_1$ and RIS $S_1$ on the $m$th subcarrier is denoted by $\mathbf{h}_{m,\rm I2} \in \mathbb{C}^{N \times 1}$ and that between receive antenna $R_1$ and RIS $S_1$ on the $m$th subcarrier is denoted by $\mathbf{h}_{m,\rm 1I} \in \mathbb{C}^{N \times 1}$. 
%We let $\mathbf{h}_{m,\rm I2} \in \mathbb{C}^{N \times 1}$ denote the frequency-domain channel between transmit antenna $T_1$ and RIS $S_1$ on the $m$th subcarrier and let $\mathbf{h}_{m,\rm 1I} \in \mathbb{C}^{N \times 1}$ denote the frequency-domain channel between receive antenna $R_1$ and RIS $S_1$ on the $m$th subcarrier.
Similarly, the frequency-domain channel between transmit antenna $T_2$ and RIS $S_2$ on the $m$th subcarrier is denoted by  $\mathbf{g}_{m,\rm I2} \in \mathbb{C}^{N \times 1}$ and that between receive antenna $R_2$ and RIS $S_2$ on the $m$th subcarrier is denoted by $\mathbf{g}_{m,\rm 1I} \in \mathbb{C}^{N \times 1}$.
%Similarly, we let $\mathbf{g}_{m,\rm I2} \in \mathbb{C}^{N \times 1}$ denote the frequency-domain channel between transmit antenna $T_2$ and RIS $S_2$ on the $m$th subcarrier and let $\mathbf{g}_{m,\rm 1I} \in \mathbb{C}^{N \times 1}$ denote the frequency-domain channel between receive antenna $R_2$ and RIS $S_2$ on the $m$th subcarrier. 
In this paper, we assume that both RISs only reflect once since the power
of the signal reflected two or more times is much smaller than
that of the signal reflected once \cite{you2021energy}, \cite{guo2019weighted}.
%Moreover, the channel gain of $T_1$-$S_2$-$R_1$ link and $T_2$-$S_1$-$R_2$ link can be neglected due to longer distance and its configuration for supporting the reception at $U_2$. 
Moreover, the channel gains of $T_1$-$S_2$-$R_1$ link and $T_2$-$S_1$-$R_2$ link can be neglected due to the long distance \cite{shen2020beamformig}, \cite{deshpande2022spatially}. 
In this paper, we assume that wireless channels experience quasi-static block-fading and the channel state information (CSI) is known to FD \cite{deshpande2022spatially,hua2022joint}. %This means that the CSI remains unchanged in the coherent time block and changes in different coherent time blocks \cite{zhang2020capacity}.

\begin{figure}[htbp]
	\centering
	\vspace{3pt}
	\includegraphics[scale=0.50]{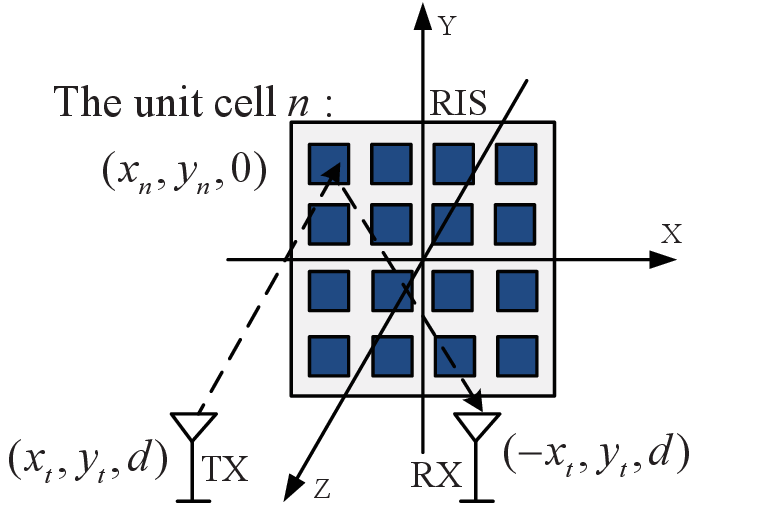}
	\caption{Right-angle coordinate system for each FD device.} \label{fig:compute_distance}
	\vspace{-1pt}
\end{figure}

 We set up a right-angle coordinate system for each FD device to compute the coordinate of the center of each unit cell and analyze the near-field channel of RIS. As shown in Fig. \ref{fig:compute_distance}, the center of RIS is located at the coordinate origin of the XY-plane. Also, both the transmit antenna and the receive antenna are located in the XZ-plane and they are symmetric about the Z axis. The coordinates of the transmit and receive antenna are denoted by $\left(x_{t}, y_{t}, d\right)$ and $\left(-x_{t}, y_{t}, d\right)$, respectively. We consider the RIS is square where the cells are deployed edge-to-edge and each RIS unit cell has an area of $R_m$ on the $m$th subcarrier. If we number the unit cells from left to right, row by row, the coordinate of the center of the $n$th unit cell on the $m$th subcarrier is $\left(x_{m,n}, y_{m,n}, 0\right)$ with  
\begin{align}
 	x_{m,n}=-\frac{(\sqrt{N}-1) \sqrt{R_m}}{2}+\sqrt{R_m} \mathrm {mod} (n-1, \sqrt{N}) 
\end{align}
and
\begin{align}
y_{m,n}=\frac{(\sqrt{N}-1) \sqrt{R_m}}{2}-\sqrt{R_m}\left\lfloor \frac{n-1}{\sqrt{N}}  \right \rfloor
\label{eq_coordinate}.
 \end{align}
%where $\bmod (\cdot , \cdot )$ denotes the modulo operation and $\left\lfloor \cdot   \right \rfloor$ denotes the operation of round down. 
We denote by $\mathbf{Q}_m \in \{ {\it h}_{m,1},{\it g}_{m,1},\mathbf{g}_{m,\rm I2},\mathbf{g}_{m,\rm 1I},\mathbf{h}_{m,\rm I2},\mathbf{h}_{m,\rm 1I}\}$ the near-field channel on the $m$th subcarrier where $\mathit m\in \mathcal{M}$ and use an exact model, which takes into account the near-field behavior of the RIS, to quantify it \cite{bjornson2020power}, \cite{dardari2020communicating}. We consider that the transmit antenna located at $\bf{t}=(\it x_t,\it y_t,\it d)$ is lossless isotropic and transmits a signal that has polarization in the Y direction when traveling in the Z direction.
%a lossless isotropic antenna located at $\bf {t} = (\it x_t,\it y_t,\it d)$ that transmits a signal that has polarization in the Y direction when traveling in the Z direction. 
The channel amplitude gain on the $m$th subcarrier $\xi_{m,\mathbf t,\mathbf{p}_{m,n}}$ at the $n$th unit cell , located at $\mathbf{p}_{m,n} = (x_{m,n}, y_{m,n}, 0)$, can be given by~\eqref{eq_channel} at bottom of this page, where 

\setcounter{equation}{3} 
\begin{align}\label{eq4}
\mathcal{X}_{m,t, n}&=\left\{\sqrt{R_m} / 2+x_{m,n}-x_{t}, \sqrt{R_m} / 2-x_{m,n}+x_{t}\right\}
\end{align}
and
\begin{align}\label{eq5}
\mathcal{Y}_{m,t, n}&=\left\{\sqrt{R_m} / 2+y_{m,n}-y_{t}, \sqrt{R_m} / 2-y_{m,n}+y_{t}\right\}.
\end{align}

\par Based on~\eqref{eq_channel},~\eqref{eq4}, and~\eqref{eq5}, the channel gain on the $m$th subcarrier between the transmit antenna and the RIS %denoted by $\boldsymbol{\varsigma}$, 
can be given by 
\begin{align}
\mathbf{Q}_m= \left[\it {\varsigma}_{m,\rm 1}, \cdots, \it {\varsigma}_{m,N}\right]^{\mathrm{T}},
\end{align}
where $\mathbf{Q}_m \in \{ \mathbf{g}_{m,\rm I2},\mathbf{g}_{m,\rm 1I},\mathbf{h}_{m,\rm I2},\mathbf{h}_{m,\rm 1I}\}$ and $
\varsigma_{m,n}=\left | \varsigma_{m,n} \right |e^{-j \vartheta_{m,n}}$
with 
\begin{align}
{\left |\varsigma_{m,n}\right|}^2&=\xi_{m,\mathbf{t}, \mathbf{p}_{m,n}}
\end{align}
and
\begin{align}\label{eq8}
\vartheta_{m,n}&=2 \pi \bmod \left(\frac{\left \| \mathbf {t}-\mathbf{p}_{m,n} \right \| }{\lambda_m}, 1\right).
\end{align}
In \eqref{eq8}, $\lambda_m=c/f_m$ is the wavelength of the $m$-th subcarrier with the frequency of 
\begin{align}
f_m=f-B/2+m\Delta f.
\end{align}
%is the phase shift of the $n$th unit cell of the RIS.
Similarly, we can also obtain the channel gain on the $m$th subcarrier between the RIS and the receive antenna using the above model. % Because the coordinates of the transmit antenna and receive antenna are fixed and the distance between them is short, the SI channel gain is fixed under different subcarriers and there is no multiple paths. 
%Due to the coordinates of the transmit antenna and receive antenna are fixed, the modulus of SI channel gain is fixed under different subcarriers.
Since the transmit and receive antenna coordinates are fixed, the modulus of the SI channel gain remains constant under different subcarriers. However, the phase of the SI channel gain varies under different subcarriers. The expression for the near-field channel gain of RIS provided in this paper is derived by treating each RIS unit cell as a receive antenna~\cite{bjornson2020power, dardari2020communicating}. Thus, when calculating the SI gain between the transmit and receive antenna, we can consider the receive antenna as a RIS which has only one unit cell.
Following similar steps, the SI channel gain on the $m$th subcarrier between the transmit antenna and the receive antenna of each FD device can be expressed as follows:
\begin{align}\label{eq10}
\mathbf{Q}_m=\sqrt{\xi_{\mathbf t,\mathbf{p}_{\rm r}}} e^{-j \vartheta_{m,\rm r}},
\end{align}
where $\mathbf{Q}_m \in \{ {\it h}_{m,1},{\it g}_{m,1} \}$. The coefficients $\xi_{\mathbf t,\mathbf{p}_{\rm r}}$ and $\vartheta_{m,\rm r}$ in~\eqref{eq10} can be equivalently calculated based on~\eqref{eq_channel},~\eqref{eq4},~\eqref{eq5} and~\eqref{eq8} with $\mathbf{p}_{\rm r} = (x_t, y_t, 0)$ and $d=2x_t$.
%When channel $\mathbf{H} \in \{\it {h}_ {\rm 1},\it {g}_{\rm 1}\}$, i.e. the channel gain between the transmit and receive antenna of each FD device, %we set $n=1$, and the coordinate of piont  $\mathbf{p}_1 = (x_1, y_1, 0)$ denote the coordinate of antenna.
%we set the coordinate of point  $\mathbf{p} = (x_t, y_t, 0)$ with $d=2x_t$, which indicates the coordinate of the receive antenna.
 %Lemma 1  provides a general way of computing channel gains to each of the $N$ unit cells of the RIS in the so-called geometric near-field of the array. 
 Note that the employed channel model is applicable to the radiating near-field, but not to the reactive near-field~\cite{bjornson2020power, dardari2020communicating}. We assume that the transmitter/receiver of each FD device is in the radiating near-field of each unit cell of its equipped RIS, such that the channel model is accurate.%. In fact, this assumption must be made to derive the expression in Lemma 1\cite{dardari2020communicating}. 
\par The far-field channel on the $m$th subcarrier is denoted by $\mathbf{\Gamma }_m \in \{\mathit{h}_{m,\rm T1},\mathit{h}_{m,\rm T2},\mathbf{g}_{m,12},\mathbf{g}_{m,21},\mathbf{h}_{m,12},\mathbf{h}_{m,21}\}$, then we model it with Rician fading. For a Rician channel with K-factor denoted by $K_{\rm r}$, the channel matrix $\boldsymbol{\rm \Upsilon  }$ can be given as follows:
\begin{align}
\mathbf{\Upsilon}=\sqrt{L}\left(\sqrt{\frac{K_{\rm r}}{1+K_{\rm r}}}\boldsymbol{\rm \Upsilon  }^{\rm LOS}+\sqrt{\frac{1}{1+K_{\rm r}}}\boldsymbol{\rm \Upsilon  }^{\rm NLOS}\right),
\end{align}
where $\boldsymbol{\rm \Upsilon  }^{\rm NLOS}$ denotes a large number of secondary non-light of sight (NLOS) channel gains, $\boldsymbol{\rm \Upsilon  }^{\rm LOS}$ denotes the normalized complex light of sight (LOS) channel gains between the transceiver antennas, and $L$ denotes the corresponding pathloss. We can obtain the far-field channel on the $m$th subcarrier $\mathbf{\Gamma }_m$ from the following equation
\begin{align}
\mathbf{\Gamma }_m= \mathbf{\Upsilon}\boldsymbol{f}_{m}^{\mathrm{H}},
\end{align}
where $\boldsymbol{f}_{m}^{\mathrm{H}}$ denotes the $m$th column of
the $M \times M$ discrete Fourier transform (DFT) matrix $\boldsymbol{F}_{M}$.

\subsection{Signal Model}
Without loss of generality, in the following we only analyze the transmission
from $U_2$ to $U_1$ with the assistance of $S_1$ and $S_2$. The presented analyses and scheme can be extended for the
transmission from $U_1$ to $U_2$ with the assistance of $S_1$ and $S_2$ in a straightforward
manner.
We denote  by $\mathbf{x}_1  \triangleq\left[X_{1,0},\ldots, X_{1,m}, \ldots, X_{1,M-1}\right]^{T}$ and $\mathbf{x}_2  \triangleq\left[X_{2,0},\ldots, X_{2,m}, \ldots, X_{2,M-1}\right]^{T}$the OFDM symbol transmitted from transmit antenna $T_1$ and that from transmit antenna $T_2$, respectively, where $X_{1,m}$ and $X_{2,m}$ are the OFDM symbols on the $m$th subcarrier. These OFDM symbols are first transformed into the time domain via an $M$-point inverse discrete Fourier transform (IDFT), and then appended by a cyclic prefix (CP) of length $M_{\rm cp}$. We assume that the length of CP is longer than the maximum delay spread of all channels, denoted by $L_{\rm max}$, to eliminate the inter-symbol interference (ISI). Moreover,
let $ \mathbf{p}_1= \left[p_{1,0}, \ldots,p_{1,m}, \ldots, p_{1,M-1}\right]^{T} \in \mathbb{R}^{M \times 1}$ and $ \mathbf{p}_2= \left[p_{2,0}, \ldots,p_{2,m}, \ldots, p_{2,M-1}\right]^{T} \in \mathbb{R}^{M \times 1}$ where $p_{1,m} \ge 0$, $p_{2,m} \ge 0$ denote the power allocated to the $m$th subcarrier of $U_1$ and $U_2$, respectively. Assume the transmit power constraint at $U_1$ is $P_1$ and the transmit power constraint at $U_2$ is $P_2$. Thus the power allocation at $U_1$ and $U_2$ should satisfy $\sum_{m=0}^{M-1} p_{1,m} \leq P_1$ and $\sum_{m=0}^{M-1} p_{2,m} \leq P_2$, respectively. In fact, when we design phase shift control circuits, RIS can only bring the same phase shift on different subcarriers. Thus we model the RC matrix of RIS $S_1$ as $\bf{\Phi}=\sqrt{\eta} \rm{diag}\left(\rm{\phi}_{1}, \cdots, \phi_\mathit{n}, \cdots, {\phi}_\mathit{N}\right)$,   where $\eta\le 1$ is the reflection efficiency and $\mathrm{\phi}_\mathit{n}$ with $\mathit n\in \mathcal{N} \triangleq\{1,2, \ldots, N\}$ is the RC of the $n$th unit cell of RIS $S_1$. In the same way, we model the RC matrix of RIS $S_2$  as $\bf{\Psi}=\sqrt{\beta } \rm{diag}\left( \psi _{1}, \cdots,  \psi_\mathit{n}, \cdots, \rm \psi _\mathit{N}\right)$, where $\beta \le 1$ is the reflection efficiency and $\rm \psi_\mathit{n}$ with $\mathit n\in \mathcal{N}$ is the RC of the $n$th unit cell of RIS $S_2$. 
Considering the practical implementation of the RIS, we assume that the feasible set of RC $\rm {\varphi}_\mathit{n}\in \{ \phi_\mathit{n},  \psi_\mathit{n} \}$   has the following three cases:

%we model the RC matrix of RIS $S_1$ as $\bf{\Phi}=\sqrt{\eta} \rm{diag}\left(\rm{\phi}_{1}, \cdots, \phi_\mathit{n}, \cdots, {\phi}_\mathit{N}\right)$, where $\eta\le 1$ is the reflection efficiency and $\mathrm{\phi}_\mathit{n}$ is the RC of the $n$th unit cell of RIS $S_1$. In the same way, we model the RC matrix of RIS $S_2$  as $\bf{\Psi}=\sqrt{\beta } \rm{diag}\left( \psi _{1}, \cdots,  \psi_\mathit{n}, \cdots, \rm \psi _\mathit{N}\right)$, where $\beta \le 1$ is the reflection efficiency and $\rm \psi_\mathit{n}$ is the RC of the $n$th unit cell of RIS $S_2$. Considering the practical implementation of the RIS, we assume that the feasible set of RC $\rm {\varphi}_\mathit{n}\in \{ \phi_\mathit{n},  \psi_\mathit{n} \}$   has the following three cases:

\begin{itemize}
	\item[1)]
	Both the amplitude and phase of each RIS unit cell can be controlled independently and continuously \cite{guo2019weighted,tang2020mimo}, i.e.,
	\begin{align}
	\rm \varphi _\mathit{n} \in \mathcal{F}_{1} \triangleq\left\{\rm \varphi _\mathit{n}\Big|\big| \rm \varphi _\mathit{n}\big|^{2} \leq 1\right\},
	\end{align}
	so the theoretical upper bound of passive beamforming performance can be achieved under this assumption.
\end{itemize}
 %\par 1) Both the amplitude and phase of each RIS unit cell can be controlled independently and continuously \cite{guo2019weighted,tang2020mimo}, i.e.
%\begin{align}
 %\rm \varphi _\mathit{n} \in \mathcal{F}_{1} \triangleq\left\{\rm \varphi _\mathit{n}\Big|\big| \rm \varphi _\mathit{n}\big|^{2} \leq 1\right\},
%\end{align}
% so the theoretical upper bound of passive beamforming performance can be achieved under this assumption.
\begin{itemize}
	\item[2)]
     The phase of each RIS unit cell can be controlled independently, while the amplitude is fixed to one \cite{guo2019weighted},\cite{wu2018intelligent}, i.e.,
     \begin{align}
     \varphi_\mathit{n} \in \mathcal{F}_{2} \triangleq\left\{ \varphi_\mathit{n}\Big|\big|  \varphi_\mathit{n}\big|^{2}=1\right\}.
    \end{align}
\end{itemize}
\begin{itemize}
	\item[3)]
    The RC of each RIS unit cell can only take $\tau$ discrete values and these discrete values are  equally spaced on the unit circle \cite{an2022low},\cite{sun2022performance},  i.e.,
    \begin{multline}
    \rm \varphi_\mathit{n} \in \mathcal{F}_{3} \triangleq\Bigg\{\rm \varphi_\mathit{n}  \Bigg| \rm \varphi_\mathit{n}=e^{j \rm \theta_\mathit{n}},\\
    \theta_{n} \in\left\{0, \frac{2 \pi}{\tau}, \cdots, \frac{2 \pi(\tau-1)}{\tau}\right\}\Bigg\}. 
    \end{multline}
\end{itemize}
By analyzing the three cases of the RIS, we can comprehensively understand the characteristics and performances of RIS-assisted FD-OFDM communication system~\cite{guo2019weighted}. The continuous and discrete phases case have lower manufacturing complexity compared to the ideal phases case which requires simultaneous adjustment of both phase and amplitude. This can ultimately provide a better guidance and support for practical applications.
\par After removing CP and performing $M$-point DFT, the received signal on the $m$th subcarrier at the receiver of device $U_1$ is expressed as follows:
\begin{multline}\label{eq15}
Y_{1,m}=\left(\mathit{h}_{m,\rm T2}+{\mathbf{h}_{m,\rm 1I}^{\mathrm{H}}} \boldsymbol{\Phi} \mathbf{h}_{m,21}+{\mathbf{g}_{m,21}^{\mathrm{H}}} \boldsymbol{\Psi} \mathbf{g}_{m,\rm I2}\right)X_{2,m} \\ +\left(\mathit{h}_{m,1}+{\mathbf{h}_{m,\rm 1I}^{\mathrm{H}} } \boldsymbol{\Phi} \mathbf{h}_{m,\rm I2}\right) X_{1,m}+\mathit{Z}_{1,m},
\end{multline}
where $\mathit{Z}_{1,m}\sim \mathcal{CN}(0,\sigma_{1}^{2})$ is  the additive white Gaussian noise (AWGN) on the $m$th subcarrier. %following $\mathcal{CN}(0,\sigma_{1}^{2})$, which indicates different subcarriers are with equal noise power.

After simplification, \eqref{eq15} can be re-written as follows:
\begin{multline}\label{eq_simply}
Y_{1,m}=\left(\mathit{h}_{m,\rm T2}+{\mathbf{h}_{m,\rm I}^{\mathrm{H}}} \boldsymbol{\phi} +{\mathbf{g}_{m,\rm I}^{\mathrm{H}}} \boldsymbol{\psi} \right)X_{2,m} \\ +\left(\mathit{h}_{m,1}+{\mathbf{h}_{m,\rm R}^{\mathrm{H}} } \boldsymbol{\phi} \right) X_{1,m}+\mathit{Z}_{1,m},
\end{multline}
 where $\mathbf{h}_{m,\rm I}=\sqrt{\eta}\left(\mathbf{h}_{m,\rm 1I} \odot \mathbf{h}_{m,\rm 21}^{*}\right)$ with $\mathbf{h}_{m,\rm I} \in \mathbb{C}^{N \times 1}$, $\mathbf{g}_{m,\rm I}=\sqrt{\beta}\left(\mathbf{g}_{m,\rm 21} \odot \mathbf{g}_{m,\rm I2}^{*}\right)$ with $\mathbf{g}_{m,\rm I} \in \mathbb{C}^{N \times 1}$, and $\mathbf{h}_{m,\rm R}=\sqrt{\eta}\left(\mathbf{h}_{m,\rm 1I} \odot \mathbf{h}_{m,\rm I2}^{*}\right)$ with $\mathbf{h}_{m,\rm R} \in \mathbb{C}^{N \times 1}$. Moreover, $\boldsymbol{\phi}=\left[\phi_{1}, \cdots, \phi_{n}, \cdots, \phi_{N}\right]^{\mathrm{T}}$ and $\boldsymbol{\psi}=\left[\psi_{1}, \cdots, \psi_{n} \cdots, \psi_{N}\right]^{\mathrm{T}}$ denote the RC vector of RIS $S_1$ and RIS $S_2$, respectively. We define $\mathbf{y}_1 \triangleq\left[Y_{1,0}, \cdots, Y_{1,m}, \cdots, Y_{1,M-1}\right]^{\mathrm{T}}$ and the received OFDM symbol at the receiver of device $U_1$ can be re-written as follows:
 \begin{multline}\label{eq_frequency}
 \mathbf{y}_1=\mathbf{X}_{2} \left(\mathbf{h}_{\rm T,2}+{\mathbf{H}_{\rm I}} \boldsymbol{\phi} +{\mathbf{G}_{\rm I}} \boldsymbol{\psi} \right) \\ +\mathbf{X}_{1}\left(\mathbf{h}_{1}+{\mathbf{H}_{\rm R}} \boldsymbol{\phi} \right) +\mathbf{z}_{1},
 \end{multline}
 where $\mathbf{h}_{\rm T,2}=\left[\mathit{h}_{0,\rm T2}, \cdots, \mathit{h}_{m,\rm T2}, \cdots, \mathit{h}_{M-1,\rm T2}\right]^{\mathrm{T}}\in \mathbb{C}^{M \times 1}$ is the channel
 frequency response (CFR) of $T_2$-$R_1$ link, $\mathbf{h}_{\rm 1}=\left[\mathit{h}_{0,\rm 1}, \cdots, \mathit{h}_{m,\rm 1}, \cdots, \mathit{h}_{M-1,\rm 1}\right]^{\mathrm{T}}\in \mathbb{C}^{M \times 1}$ is the CFR of SI channel, $\mathbf{H}_{\rm I}\boldsymbol{\phi}$ with $\mathbf{H}_{\rm I}=\left[\mathit{h}_{0,\rm I}^{\mathrm{H}}, \cdots, \mathit{h}_{m,\rm I}^{\mathrm{H}}, \cdots, \mathit{h}_{M-1,\rm I}^{\mathrm{H}}\right]^{\mathrm{T}}\in \mathbb{C}^{M \times N}$ is the equivalent
 cascaded CFR of $T_2$-$S_1$-$R_1$ link, $\mathbf{G}_{\rm I} \boldsymbol{\psi}$ with  $\mathbf{G}_{\rm I}=\left[\mathit{g}_{0,\rm I}^{\mathrm{H}}, \cdots, \mathit{g}_{m,\rm I}^{\mathrm{H}}, \cdots, \mathit{g}_{M-1,\rm I}^{\mathrm{H}}\right]^{\mathrm{T}}\in \mathbb{C}^{M \times N}$ is the equivalent
 cascaded CFR of $T_2$-$S_2$-$R_1$ link, and ${\mathbf{H}_{\rm R}} \boldsymbol{\phi}$ with  $\mathbf{H}_{\rm R}=\left[\mathit{h}_{0,\rm R}^{\mathrm{H}}, \cdots, \mathit{h}_{m,\rm R}^{\mathrm{H}}, \cdots, \mathit{h}_{M-1,\rm I}^{\mathrm{H}}\right]^{\mathrm{T}}\in \mathbb{C}^{M \times N}$ is the equivalent
 cascaded CFR of $T_1$-$S_1$-$R_1$ link. $\mathbf{X}_{2}=\rm{diag}\left(\mathbf{x}_2\right)$ and $\mathbf{X}_{1}=\rm{diag}\left(\mathbf{x}_1\right)$ denote the diagonal matrix of OFDM
 symbol $\mathbf{x}_2$ and OFDM
 symbol $\mathbf{x}_1$, respectively. Moreover,  $\mathbf{z}_1 =\left[Z_{1,0}, \cdots, Z_{1,m}, \cdots, Z_{1,M-1}\right]^{\mathrm{T}}$ denotes the AWGN vector at $U_1$, which follows $\mathcal{CN}(0,\sigma_{1}^{2}\rm {I}_\mathit {M})$. In addition, $\mathbf{X}_{2} \left(\mathbf{h}_{\rm T,2}+{\mathbf{H}_{\rm I}} \boldsymbol{\phi} +{\mathbf{G}_{\rm I}} \boldsymbol{\psi} \right)$ represents the desired signal, $\mathbf{X}_{1}\mathbf{h}_{1}$ is the SI, and $\mathbf{X}_{1}{\mathbf{H}_{\rm R}} \boldsymbol{\phi}$ is the RIS-assisted signal to cancel the severe SI. The SIC performance can be measured by the SIC capability $\Lambda $ (dB) \cite{cheng2019achieving}, which is defined as follows:
\begin{align}\label{eq16}
\Lambda=10 \log _{10} \frac{E_{0}+E_{N}}{E_{c}+E_{N}},
\end{align}
where $E_{0}$ is the energy of the SI signal before SIC, $E_{c}$
is the energy of the residual SI after SIC, and $E_{N}$ is the energy of noise.
\par Substituting Eq.~\eqref{eq_simply} into Eq.~\eqref{eq16}, we have the SIC capability of RIS-assisted FD-OFDM communications as follows:
\begin{align}\label{SIC capability}
\Lambda=10 \log _{10} \sum_{m=0}^{M-1}{\frac{\left|\mathit{h}_{m,1}\right|^{2} p_{1,m}+\sigma_{1}^{2}}{\left|\mathit{h}_{m,1}+{\mathbf{h}_{m,\rm R}^{\mathrm{H}} } \boldsymbol{\phi}\right|^{2} p_{1,m}+\sigma_{1}^{2}}} .
\end{align}
Furthermore, the system capacity after SIC can be expressed as follows:
\begin{multline} \label{Capacity_FD}
C_{\rm FD}=\frac{1}{M+M_{cp}} \sum_{m=0}^{M-1} \mathrm{\log} _{2}\left(1+\right.\\
\left.\frac{\left|\mathit{h}_{m,\rm T2}+{\mathbf{h}_{m,\rm I}^{\mathrm{H}}} \boldsymbol{\phi} +{\mathbf{g}_{m,\rm I}^{\mathrm{H}}} \boldsymbol{\psi}\right|^{2} p_{2,m}}{\left|\mathit{h}_{m,1}+{\mathbf{h}_{m,\rm R}^{\mathrm{H}} } \boldsymbol{\phi}\right|^{2} p_{1,m}+\sigma_{1}^{2}}\right).
\end{multline}
Given the same bandwidth, the capacity of HD-OFDM communication system without RIS-based SIC can be given by
\begin{align}
C_{\rm HD}=\frac{1}{2} \frac{1}{M+M_{cp}} \sum_{m=0}^{M-1} \log _{2}\left(1+\frac{\left|\mathit{h}_{m,\rm T2}\right|^{2} p_{2,m}}{\sigma_{1}^{2}}\right).
\end{align}

\section{SIC Capability Maximization Via Joint Power Allocation And Reflection Coefficient Optimization}\label {sec:Residual SI Minimization}
In this section, the RC matrix of the RIS and power allocation scheme of each device are jointly optimized to maximize the SIC capability shown in \eqref{SIC capability}. We first formulate
the optimization problem and then propose an AO algorithm to solve it. 
\subsection{Problem Formulation}
%We only show the optimization steps for RC matrix $\boldsymbol{\Phi}$. RC matrix $\boldsymbol{\Psi}$ can be optimized with the same manner. 
% In this section we only optimize the phase shift matrix $\boldsymbol{\Phi}$, because the phase shift matrix $\boldsymbol{\Psi}$ can be optimized in the same way. 
%In order to make the capacity of FD double the capacity of HD, the SIC capability should be greatly improved after RISs are deployed, which means that the component of residual SI $\left|\mathit{h}_{\mathrm{1}}+\mathbf{h}_{\mathrm{1I}}^{\mathrm{H}}  \boldsymbol{\Phi} \mathbf{h}_{\mathrm{I2}}\right|^{2}$ minimization. Meanwhile, the RIS reflection coefficients subject to the feasible set constraints,
Considering the constraints of the transmit power and the RC, the SIC capability maximization problem is formulated as follows:
\begin{subequations}
\begin{align}
	\text { (P1) }\quad  \max _{\mathbf p_{1}, \boldsymbol{\phi}} \quad&\sum_{m=0}^{M-1}{\frac{\left|\mathit{h}_{m,1}\right|^{2} p_{1,m}+\sigma_{1}^{2}}{\left|\mathit{h}_{m,1}+{\mathbf{h}_{m,\rm R}^{\mathrm{H}} } \boldsymbol{\phi}\right|^{2} p_{1,m}+\sigma_{1}^{2}}}\label{problema}\\
	\text { s.t.:} \quad  &1). \quad\sum_{m=0}^{M-1} p_{1,m} \leq P_1,\label{problemb}\\\quad  &2).\quad p_{1,m}\ge 0, \quad \forall \mathit m\in \mathcal{M},\label{problemc}\\\quad  &3). \quad\phi_{n} \in \mathcal{F}, \quad \forall \mathit n\in \mathcal{N},\label{problemd}
\end{align}
\end{subequations}
where $\mathcal{F}$ is chosen from $\mathcal{F}_1$, $\mathcal{F}_2$, and $\mathcal{F}_3$. %We can simplify $\left|\mathit{h}_{\mathrm{1}}+\mathbf{h}_{\mathrm{1I}}^{\mathrm{H}}  \boldsymbol{\Phi} \mathbf{h}_{\mathrm{I2}}\right|^{2}$ as follows:
%\begin{align}
 %\left|\mathit{h}_{\mathrm{1}}+\mathbf{h}_{\mathrm{1I}}^{\mathrm{H}}  \boldsymbol{\Phi} \mathbf{h}_{\mathrm{I2}}\right|^{2}=\left|\mathit{h}_{1}+\mathbf{h}_{\mathrm{I}}^{\mathrm{H}} \boldsymbol{\rm \phi}\right|^{2},
%\end{align}
 % where $\boldsymbol{\phi}=\left[\phi_{1}, \cdots, \phi_{N}\right]^{\mathrm{T}}$, $\mathbf{h}_\mathrm{I}=\sqrt{\eta}\left(\mathbf{h}_\mathrm{1I} \odot \mathbf{h}_\mathrm{I2}^{*}\right)$ with $\mathbf{h}_\mathrm{I} \in \mathbb{C}^{N \times 1}$, and  $\odot$ denotes Hadamard product. Then, the optimization problem can be transformed into the following problem:
%\begin{equation}\label{minP1'}
%\begin{aligned}
%\text { (P1$'$) }\quad  \min _{\boldsymbol{\phi}} \quad&\left|\mathit{h}_{1}+\mathbf{h}_{\mathrm{I}}^{\mathrm{H}} \boldsymbol{\phi}\right|^{2} \\
%\text { s.t.} \quad  &\mathrm{\phi}_{n} \in \mathcal{F}, \quad \forall n=1, \ldots, N,
%%\end{equation}
%Both constraint \eqref{problemd}  with $\mathrm{\phi}_{n} \in \mathcal{F}_{2}$ and $\mathrm{\phi}_{n} \in \mathcal{F}_{3}$ and the objective function are non-convex. 
If $\mathrm{\phi}_{n} \in \mathcal{F}_{2}$ or  $\mathrm{\phi}_{n} \in \mathcal{F}_{3}$, constraint \eqref{problemd} is non-convex. Moreover, the objective function of (P1) is non-convex.
Thus, (P1) is a non-convex problem and is difficult to solve directly. To overcome the above challenges, in the following we propose an AO algorithm which optimizes $\mathbf p_{1}$ and $\boldsymbol{\phi}$ iteratively to find an
approximate solution to (P1). %by iteratively optimizing one of $\mathbf p_{1}$ and $\boldsymbol{\phi}$ with the other fixed at each time.
\subsection{Power Allocation Optimization with Given Reflection Coefficient}
In this subsection, we optimize power allocation vector for the given RC vector. Due to the fixed RC vector, the superimposed CFR $\mathit{h}_{m,1}+{\mathbf{h}_{m,\rm R}^{\mathrm{H}} } \boldsymbol{\phi}$ is fixed, then problem (P1) can be transformed into 
\begin{subequations}
\begin{align}
\text { (P2) }\quad  \max _{\mathbf p_{1}} \quad&\sum_{m=0}^{M-1}{\frac{\mathit{b}_{m} p_{1,m}+\sigma_{1}^{2}}{\mathit{v}_{m}  p_{1,m}+\sigma_{1}^{2}}}\label{p2a}\\
\text { s.t.:} \quad  &1). \quad\sum_{m=0}^{M-1} p_{1,m} \leq P_1,\label{p2b}\\\quad  &2).\quad p_{1,m}\ge 0, \quad \forall \mathit m\in \mathcal{M},\label{p2c}
\end{align}
\end{subequations}
where $\mathit{b}_{m}=\left|\mathit{h}_{m,1}\right|^{2}$, $\mathit{v}_{m}=\left|\mathit{h}_{m,1}+{\mathbf{h}_{m,\rm R}^{\mathrm{H}} } \boldsymbol{\phi}\right|^{2}$, and constraint \eqref{problemd} is omitted because $\boldsymbol{\phi}$ is fixed. Note that (P2) is a multiple-ratio fractional programming (FP) problem which is difficult to solve, thus we use the quadratic transform \cite{shen2018fractional} to transform (P2) into the following problem 
\begin{subequations}
	\begin{align}
	\text { (P3) }  \max _{\mathbf p_{1},\mathbf q} &\sum_{m=0}^{M-1}\left({2\mathit{q}_{m}\sqrt{\mathit{b}_{m} p_{1,m}+\sigma_{1}^{2}}-\mathit{q}_{m}^{2}\left(\mathit{v}_{m}  p_{1,m}+\sigma_{1}^{2}\right)}\right)\label{p3a}\\
	\quad\text { s.t.:} \quad  &1). \quad\sum_{m=0}^{M-1} p_{1,m} \leq P_1,\label{p3b}\\\quad  &2).\quad p_{1,m}\ge 0, \quad \forall \mathit m\in \mathcal{M},\label{p3c}\\\quad  &3).\quad q_{m}\in \mathbb{R}, \quad \forall \mathit m\in \mathcal{M},\label{p3d}
	\end{align}
\end{subequations}
where $\mathbf{q}=\left[\mathit{q}_{0}, \cdots, \mathit{q}_{m}, \cdots, \mathit{q}_{M-1}\right]^{\mathrm{T}}$ includes the introduced auxiliary variables $\mathit{q}_{0}, \cdots, \mathit{q}_{m}, \cdots, \mathit{q}_{M-1}$. When the power allocation vector $\mathbf p_{1}$ is fixed, the closed-form solution to $\mathbf{q}$ can be derived as follows: 
\begin{align}\label{25}
\ddot{\mathit{q}}_{m}=\frac{\sqrt{\mathit{b}_{m} p_{1,m}+\sigma_{1}^{2}}}{\mathit{v}_{m}  p_{1,m}+\sigma_{1}^{2}},\quad \forall \mathit m\in \mathcal{M}.
\end{align}
When $\mathbf{q}$ is fixed, (P3) can be simplified to
\begin{subequations}
	\begin{align}
	\text { (P3$'$) } \! \max _{\mathbf p_{1}} &\!\sum_{m=0}^{M-1}\!\left({2\mathit{q}_{m}\sqrt{\mathit{b}_{m} p_{1,m}\!+\sigma_{1}^{2}}-\mathit{q}_{m}^{2}\left(\mathit{v}_{m}  p_{1,m}\!+\sigma_{1}^{2}\right)}\right)\label{p3`a}\\
	\quad\text { s.t.:} \quad  &1). \quad\sum_{m=0}^{M-1} p_{1,m} \leq P_1,\label{p3`b}\\\quad  &2).\quad p_{1,m}\ge 0, \quad \forall \mathit m\in \mathcal{M}.\label{p3`c}
	\end{align}
\end{subequations}
Notice that $\eqref{p3`a}$ is concave with respect to $\mathbf p_{1}$, thus (P3$'$) is a convex problem which can be solved with KKT conditions\cite{boyd2004convex}. We can write the Lagrangian function corresponding to (P3$'$) as follows:
\begin{multline}
L(\boldsymbol{\mathbf p_{1}},\lambda_{1}, \boldsymbol{\lambda}_2)=F(\mathbf p_{1})+\lambda_{1}\left(P_1-\sum_{m=0}^{M-1} p_{1,m}\right)\\+\sum_{m=0}^{M-1} \lambda_{2,m}p_{1,m},
\end{multline}
where 
\begin{align}
F(\mathbf p_{1})=\sum_{m=0}^{M-1}\left({2\mathit{q}_{m}\sqrt{\mathit{b}_{m} p_{1,m}+\sigma_{1}^{2}}-\mathit{q}_{m}^{2}\left(\mathit{v}_{m}  p_{1,m}+\sigma_{1}^{2}\right)}\right),
\end{align}
$\boldsymbol{\lambda}_2= \left[\lambda_{2,0}, \ldots, \lambda_{2,M-1}\right]$ with $\lambda_{2,m}\ge0,\forall \mathit m\in \mathcal{M}$ is the Lagrange multiplier vector, and $\lambda_1\ge0$ is the Lagrange multiplier. According to KKT conditions, we have
\begin{align}\label{kkt}
\begin{cases}
 \displaystyle\frac{\mathit{q}_{m}\mathit{b}_{m}}{\sqrt{\mathit{b}_{m} p_{1,m}+\sigma_{1}^{2}}} -\mathit{q}_{m}^{2}\mathit{v}_{m} -\lambda_{1}-\lambda_{2,m}=0, \quad \forall \mathit m\in \mathcal{M},\\
\lambda_{1}\left(P_1-\sum_{m=0}^{M-1} p_{1,m}\right) = 0, \quad  \\
\lambda_{2,m}p_{1,m}=0,\quad \forall \mathit{m} \in \mathcal{M},\quad  \\
P_1-\sum_{m=0}^{M-1} p_{1,m}\ge 0,\quad  \\
p_{1,m}\ge 0, \quad \forall \mathit m\in \mathcal{M}.
\end{cases}
\end{align}
Solving \eqref{kkt}, the closed-form solution to (P3$'$) can be directly derived as follows: 
\begin{align}\label{30}
\ddot{\mathit{p}}_{1,m}=\max \left({a}_{m}-\frac{\sigma_{1}^{2}}{\it{ b}_{m}} ,0\right),\quad \forall \mathit m\in \mathcal{M},
\end{align}
where 
\begin{align}
\mathit{a}_{m}= \frac{\mathit{q}_{m}^{2}\mathit{b}_{m}}{\left(\lambda_{1}+\mathit{q}_{m}^{2}\mathit{v}_{m}\right)^{2}} ,\quad \forall \mathit m\in \mathcal{M}
\end{align}
should be chosen to satisfy $\sum_{m=0}^{M-1} p_{1,m} = P_1$. The optimal solution to $\lambda_{1}$ and $\boldsymbol{\lambda}_2$ can be found by sub-gradient
method or ellipsoid method \cite{boyd2004convex}.

\subsection{Reflection Coefficient Optimization with Given Power Allocation}
For the given multi-carrier power allocation, the
optimization of the RC in three cases is discussed in this subsection. We only show the optimization steps for RC vector $\boldsymbol{\phi}$. RC vector $\boldsymbol{\psi}$ can be optimized in the same manner.  Due to the fixed power allocation vector $\mathbf p_{1}$, (P1) can be simplified as follows:
\begin{subequations}
	\begin{align}
	\text { (P4) }\quad  \max _{ \boldsymbol{\phi}} \quad&\sum_{m=0}^{M-1}{\frac{\left|\mathit{h}_{m,1}\right|^{2} p_{1,m}+\sigma_{1}^{2}}{\left|\mathit{h}_{m,1}+{\mathbf{h}_{m,\rm R}^{\mathrm{H}} } \boldsymbol{\phi}\right|^{2} p_{1,m}+\sigma_{1}^{2}}}\label{p4a}\\
	\text { s.t.:} \quad  &1).  \quad\phi_{n} \in \mathcal{F}, \quad \forall \mathit n\in \mathcal{N},\label{p4b}
	\end{align}
\end{subequations}
 where constraints \eqref{problemb} and \eqref{problemc} are omitted because of the fixed power allocation vector $\mathbf p_{1}$. Notice that, similar to the previous subsection, (P4) is a multiple-ratio FP problem. Similarly, we introduce $\mathbf{l}=\left[\mathit{l}_{0}, \cdots, \mathit{l}_{m}, \cdots, \mathit{l}_{M-1}\right]^{\mathrm{T}}$, which includes the introduced auxiliary variables $\mathit{l}_{0}, \cdots, \mathit{l}_{m}, \cdots, \mathit{l}_{M-1}$ and then use the quadratic transform to transform (P4) into the following problem 
\begin{subequations}
 	\begin{align}
 	\text { (P5) } \quad   \max _{\boldsymbol{\phi},\mathbf l} \quad  &\mathit{g}\left(\boldsymbol{\phi},\mathbf l\right)\label{p5a}\\
 	\quad\text { s.t.:} \quad  &1). \quad\phi_{n} \in \mathcal{F}, \quad \forall \mathit n\in \mathcal{N},\label{p5b}\\\quad  &2). \quad l_{m}\in \mathbb{R}, \quad \forall \mathit m\in \mathcal{M},\label{p5c}
 	\end{align}
\end{subequations}
where 
\begin{multline}
\mathit{g}\left(\boldsymbol{\phi},\mathbf l\right)=\sum_{m=0}^{M-1}\left(2\mathit{l}_{m}\sqrt{\left|\mathit{h}_{m,1}\right|^{2} p_{1,m}+\sigma_{1}^{2}}\right.\\ \left.-\mathit{l}_{m}^{2}\left(\left|\mathit{h}_{m,1}+{\mathbf{h}_{m,\rm R}^{\mathrm{H}} } \boldsymbol{\phi}\right|^{2} p_{1,m}+\sigma_{1}^{2}\right)\right).
\end{multline}
%$\mathbf{l}=\left[l_{0}, \cdots, l_{m}, \cdots, l_{M-1}\right]^{\mathrm{T}}$ is the auxiliary variable that we introduced. 
When $\boldsymbol{\phi}$ is fixed, the closed-form solution to $\mathbf{l}$ can be found as follows: 
\begin{align}\label{35}
\ddot{\mathit{l}}_{m}=\frac{\sqrt{\left|\mathit{h}_{m,1}\right|^{2} p_{1,m}+\sigma_{1}^{2}}}{\left|\mathit{h}_{m,1}+{\mathbf{h}_{m,\rm R}^{\mathrm{H}} } \boldsymbol{\phi}\right|^{2} p_{1,m}+\sigma_{1}^{2}},\quad \forall \mathit m\in \mathcal{M}.
\end{align}
Because of the fixed $\mathbf{l}$ and $\mathbf p_{1}$, maximizing $\mathit{g}\left(\boldsymbol{\phi},\mathbf l\right)$ is equivalent to minimizing $\sum_{m=0}^{M-1}\left(\mathit{l}_{m}^{2}\left|\mathit{h}_{m,1}+{\mathbf{h}_{m,\rm R}^{\mathrm{H}} } \boldsymbol{\phi}\right|^{2} p_{1,m} \right)$. We first omit irrelevant items and then transform (P5) into 
\begin{subequations}
\begin{align}
\text { (P6) } \quad   \min _{\boldsymbol{\phi}} \quad  &\bar{\mathit{g}}\left(\boldsymbol{\phi}\right)\label{p6a}\\
\quad\text { s.t.:} \quad  &\phi_{n} \in \mathcal{F}, \quad \forall \mathit n\in \mathcal{N},\label{p6b} 
\end{align}
\end{subequations}
where
%\begin{align}\label{g(phi)} 
%&\bar{\mathit{g}}\left(\boldsymbol{\phi}\right)\nonumber\\&\!=\!\!\!\sum_{m=0}^{M-1}\!\!\left(\!\mathit{l}_{m}^{2} p_{1,m}\!\left(\!{\mathbf{h}_{m,\rm R}^{\mathrm{H}} } \boldsymbol{\phi}\mathit{h}_{m,1}^{\ast }\!+\!\boldsymbol{\phi}^{\mathrm{H}}\mathbf{h}_{m,\rm R}\mathit{h}_{m,1}\!\!+\!\boldsymbol{\phi}^{\mathrm{H}}\mathbf{h}_{m,\rm R}\mathbf{h}_{m,\rm R}^{\mathrm{H}} \boldsymbol{\phi}\!\right)\!\!\right)\nonumber\\
%&\overset{(a)}{=}\sum_{m=0}^{M-1}\left( \mathit{w}_{m}{\mathbf{h}_{m,\rm R}^{\mathrm{H}} }\boldsymbol{\phi}+\boldsymbol{\phi}^{\mathrm{H}}\mathbf{h}_{m,\rm R}\mathit{c}_{m}\right)+\boldsymbol{\phi}^{\mathrm{H}}\mathbf{Q}\boldsymbol{\phi}\nonumber\\
%&\overset{(b)}{=}\mathbf{w}^{\mathrm{H}}\mathbf{H}_2\boldsymbol{\phi}+\boldsymbol{\phi}^{\mathrm{H}}\mathbf{H}_{2}^{\mathrm{H}}\mathbf{c}+\boldsymbol{\phi}^{\mathrm{H}}\mathbf{Q}\boldsymbol{\phi}
%\end{align}
\begin{align}\label{g(phi)} 
&\bar{\mathit{g}}\left(\boldsymbol{\phi}\right)\nonumber\\&\!=
\!\!\!\sum_{m=0}^{M-1}\!\!\left(\!\mathit{l}_{m}^{2} p_{1,m}\!\left(\!{\mathbf{h}_{m,\rm R}^{\mathrm{H}} } \boldsymbol{\phi}\mathit{h}_{m,1}^{\ast }\!+\!\boldsymbol{\phi}^{\mathrm{H}}\mathbf{h}_{m,\rm R}\mathit{h}_{m,1}\!\!+\!\boldsymbol{\phi}^{\mathrm{H}}\mathbf{h}_{m,\rm R}\mathbf{h}_{m,\rm R}^{\mathrm{H}} \boldsymbol{\phi}\!\right)\!\!\right)\nonumber\\
&=\mathbf{w}\boldsymbol{\phi}+\boldsymbol{\phi}^{\mathrm{H}}\mathbf{c}+\boldsymbol{\phi}^{\mathrm{H}}\mathbf{A}\boldsymbol{\phi}
\end{align}
with
%with 
%$\mathbf{w}=\left[\mathit{w}_{0}, \cdots, \mathit{w}_{m}, \cdots, \mathit{w}_{M-1}\right]^{\mathrm{T}}$, $\mathit{w}_{m}=\mathit{l}_{m}^{2} p_{1,m}\mathit{h}_{m,1}^{\ast }$, $\mathbf{c}=\left[\mathit{c}_{0}, \cdots, \mathit{c}_{m}, \cdots, \mathit{c}_{M-1}\right]^{\mathrm{T}}$, $\mathit{c}_{m}=\mathit{l}_{m}^{2} p_{1,m}\mathit{h}_{m,1}$, and $\mathbf{Q}=\sum_{m=0}^{M-1}\left(\mathit{l}_{m}^{2} p_{1,m}\mathbf{h}_{m,\rm R}\mathbf{h}_{m,\rm R}^{\mathrm{H}}\right)$.
\begin{align}
\mathbf{w}=\sum_{m=0}^{M-1}\left(\mathit{l}_{m}^{2} p_{1,m}\mathit{h}_{m,1}^{\ast }\mathbf{h}_{m,\rm R}^{\mathrm{H}}\right),
\end{align}
\begin{align}
\mathbf{c}=\sum_{m=0}^{M-1}\left(\mathbf{h}_{m,\rm R}\mathit{l}_{m}^{2} p_{1,m}\mathit{h}_{m,1}\right),
\end{align}
and
\begin{align}
\mathbf{A}=\sum_{m=0}^{M-1}\left(\mathit{l}_{m}^{2} p_{1,m}\mathbf{h}_{m,\rm R}\mathbf{h}_{m,\rm R}^{\mathrm{H}}\right).
\end{align}
 %$\mathbf{w}=\sum_{m=0}^{M-1}\left(\mathit{l}_{m}^{2} p_{1,m}\mathit{h}_{m,1}^{\ast }\mathbf{h}_{m,\rm R}^{\mathrm{H}}\right)$, $\mathbf{c}=\sum_{m=0}^{M-1}\left(\mathbf{h}_{m,\rm R}\mathit{l}_{m}^{2} p_{1,m}\mathit{h}_{m,1}\right)$, and $\mathbf{Q}=\sum_{m=0}^{M-1}\left(\mathit{l}_{m}^{2} p_{1,m}\mathbf{h}_{m,\rm R}\mathbf{h}_{m,\rm R}^{\mathrm{H}}\right)$.
Notice that \eqref{g(phi)} is a quadratic convex function with respect to $\boldsymbol{\phi}$. Thus, (P6) is a quadratically constrained quadratic programming (QCQP). If $\mathrm{\phi}_{n} \in \mathcal{F}_{1}$, (P6) is a convex QCQP and its closed-form solution can be directly obtained using convex optimization tools. If $\mathrm{\phi}_{n} \in \mathcal{F}_{2}$ or  $\mathrm{\phi}_{n} \in \mathcal{F}_{3}$, (P6) is a non-convex problem and we can only obtain a suboptimal solution to (P6).

 % If $\mathrm{\phi}_{n} \in \mathcal{F}_{2}$, (P1$'$) is a non-convex problem and needs to be transformed into a convex problem to solve. If $\mathrm{\phi}_{n} \in \mathcal{F}_{3}$, (P1$'$) is a non-convex problem. We can obtain the suboptimal solution by projecting the optimal solution from $\mathcal{F}_{1}$ to the nearest point of the feasible set $\mathcal{F}_{3}$.
\par 1) If $\mathrm{\phi}_{n} \in \mathcal{F}_{1}$, constraint \eqref{p6b} is  
\begin{align}
\left|\mathrm{\phi}_{n}\right|^{2} \leq 1, \quad \forall \mathit n\in \mathcal{N},
\end{align}
which can be re-written as follows:
\begin{align}
 \boldsymbol{\phi}^{\mathrm{H}} \boldsymbol{\varepsilon}_{n} \boldsymbol{\varepsilon}_{n}^{\mathrm{H}} \boldsymbol{\phi} \leq 1, \quad \forall \mathit n\in \mathcal{N},
\end{align}
where $\boldsymbol{\varepsilon}_{n} \in \mathbb{R}^{N \times 1}$ is a column vector whose elements are all zero except that the $n$th is one. Thus, (P6) can be further transformed into
\begin{equation}
\begin{aligned}
\text { (P6$'$) }\quad  \min _{\boldsymbol{\phi}} \quad&\bar{\mathit{g}}\left(\boldsymbol{\phi}\right) \\
\text { s.t.} \quad  &\boldsymbol{\phi}^{\mathrm{H}} \boldsymbol{\varepsilon}_{n} \boldsymbol{\varepsilon}_{n}^{\mathrm{H}} \boldsymbol{\phi} \leq 1, \quad \forall \mathit n\in \mathcal{N}.
\end{aligned}
\end{equation}
Note that (P6$'$) is convex and can be solved with  KKT conditions. We can write the Lagrangian function corresponding to  (P6$'$) as follows:
\begin{align}
L(\boldsymbol{\phi}, \boldsymbol{\nu})=\bar{\mathit{g}}\left(\boldsymbol{\phi}\right)+\sum_{n=1}^{N} \nu_{n}\left(\boldsymbol{\phi}^{\mathrm{H}} \boldsymbol{\varepsilon}_{n} \boldsymbol{\varepsilon}_{n}^{\mathrm{H}} \boldsymbol{\phi}-1\right),
\end{align}
where $\boldsymbol{\nu}= \left[\nu_{1}, \ldots, \nu_{N}\right]$ is the Lagrange multiplier vector with $\nu_n\ge0$ for $\forall \mathit n\in \mathcal{N}$. According to KKT conditions, we have 

\begin{align}\label{eqkkt}
\begin{cases}
2\mathbf{A}\boldsymbol{\phi}+\mathbf{c}+\mathbf{w}^{\mathrm{H}}+2\sum_{n=1}^N \nu_{n}\boldsymbol{\varepsilon}_{n} \boldsymbol{\varepsilon}_{n}^{\mathrm{H}} \boldsymbol{\phi}=0, \\
	\boldsymbol{\phi}^{\mathrm{H}} \boldsymbol{\varepsilon}_{n} \boldsymbol{\varepsilon}_{n}^{\mathrm{H}} \boldsymbol{\phi}-1 \leq 0, \quad \forall \mathit n\in \mathcal{N}, \\
	\nu_{n}\left(\boldsymbol{\phi}^{\mathrm{H}} \boldsymbol{\varepsilon}_{n} \boldsymbol{\varepsilon}_{n}^{\mathrm{H}} \boldsymbol{\phi}-1\right)=0,\quad  \forall \mathit n\in \mathcal{N}.
\end{cases}
\end{align}
Solving~\eqref{eqkkt}, we can obtain the optimal solution as follows:
\begin{align} \label{43}
\ddot{\boldsymbol{\phi}}^{\mathcal{F}_{1}}=-\frac{1}{2}\left(\sum_{n=1}^{N} \nu_{n} \boldsymbol{\varepsilon}_{n} \boldsymbol{\varepsilon}_{n}^{H}+\mathbf{A}\right)^{-1} \left(\mathbf{c}+\mathbf{w}^{\mathrm{H}}\right),
\end{align}
where the optimal $\nu_{n}$, denoted by $\ddot{\nu}_{n}$, can be obtained by sub-gradient
method or ellipsoid method.
\par 2) If $\mathrm{\phi}_{n} \in \mathcal{F}_{2}$, constraint \eqref{p6b} is  
\begin{align}
\left|\mathrm{\phi}_{n}\right|^{2}=1, \quad \forall \mathit n\in \mathcal{N},
\end{align}
which is equivalent to $\left|\mathrm{\phi}_{n}\right|=1, \forall \mathit n\in \mathcal{N}.$ %Besides, $\left|\mathit{h}_{1}+\mathbf{h}_{\mathrm{I}}^{\mathrm{H}} \boldsymbol{\phi}\right|^{2}=\mathit{h}_{1}^{\mathrm{\ast}}\mathit{h}_{1}+\mathbf{h}_{\mathrm{I}}^{\mathrm{H}} \boldsymbol{\phi}\mathit{h}_{1}^{\mathrm{\ast}}+\boldsymbol{\phi}^{\mathrm{H}} \mathbf{h}_{\mathrm{I}} \mathit{h}_{1}+\boldsymbol{\phi}^{\mathrm{H}} \mathbf{h}_{\mathrm{I}} \mathbf{h}_{\mathrm{I}}^{\mathrm{H}} \boldsymbol{\phi},$ afer we omit the irrelevant term because of the optimization of variable $\boldsymbol{\phi}$, optimization problem
Then, (P6) can be transformed into the following problem
\begin{subequations}
\begin{align}
\text { (P6$''$) }\quad  \min _{\boldsymbol{\phi}} \quad&\bar{\mathit{g}}\left(\boldsymbol{\phi}\right) \label{p6''a}\\
\text { s.t.} \quad  &\left|\phi_{n}\right|=1, \quad \forall \mathit n\in \mathcal{N}\label{p6''b}.
\end{align}
\end{subequations}
Due to the non-convex constraint \eqref{p6''b}, (P6$''$) is a non-convex QCQP. In general, the unit modulus constraint problem is solved by dropping the rank-one constraint using semidefinite relaxation (SDR) method \cite{wu2018intelligent}. However, SDR method has high computational complexity and may not converge to a good suboptimal solution. Thus, we propose RCG algorithm to solve (P6$''$).

According to \cite{li2022joint, absil2008optimization}, unit modulus constraint \eqref{p6''b} forms an $N$-dimensional complex circle manifold denoted as follows:
\begin{align}
\mathcal{M}_{cc}^{N}=\left\{\boldsymbol{\phi} \in \mathbb{C}^{N}:\left|\phi_{1}\right|=\left|\phi_{2}\right|=\cdots=\left|\phi_{N}\right|=1\right\},
\end{align}
where complex circle manifold $\mathcal{M}_{cc}^{N}$ is a Riemannian sub-manifold of Riemannian manifold $\mathbb{C}^{N}$.
Thus, (P6$''$) can be formulated as the following problem %Then, RCG algorithm can be utilized to effectively find a solution. 
\begin{equation}
\begin{aligned}
\text { (P7) }\quad  \min _{\boldsymbol{\phi} \in \mathcal{M}_{\mathbf{cc}}^{N}} \quad\bar{\mathit{g}}\left(\boldsymbol{\phi}\right), 
\end{aligned}
\end{equation}
where (P7) is an unconstrained problem on the surface of $\mathcal{M}_{cc}^{N}$.
For any point $\boldsymbol{\phi}_{k}$ on $\mathcal{M}_{cc}^{N}$, the tangent space formed by passing through this point can be represented as
\begin{align}\label{49}
\mathcal{T}  _{\boldsymbol{\phi}_{k}} \mathcal{M}_{cc}^{N}=\left\{\boldsymbol{\mu} \in \mathbb{C}^{N}: \Re\left(\boldsymbol{\mu} \odot \boldsymbol{\phi}_{k}^{*}=\mathbf{0}_{N}\right)\right\},
\end{align}
where $\boldsymbol{\mu}$ is a tangent vector passing through $\boldsymbol{\phi}_{k}$. 

Riemannian gradient \cite{absil2008optimization} is defined as the tangent vector or direction in the Riemannian tangent space, which represents the direction of the fastest change of the objective function. Since $\mathcal{M}_{cc}^{N}$ is a Riemannian sub-manifold, the Riemannian gradient of $\bar{\mathit{g}}\left(\boldsymbol{\phi}_k\right)$ at point $\boldsymbol{\phi}_{k}$, denoted by $\nabla_\mathcal{M} \bar{\mathit{g}}\left(\boldsymbol{\phi}_{k}\right)$, can be expressed as follows:
\begin{align}\label{50}
\nabla_{\mathcal{M}} \bar{\mathit{g}}\left(\boldsymbol{\phi}_{k}\right)&=\operatorname{Proj}\left ( \nabla \bar{\mathit{g}}\left(\boldsymbol{\phi}_{k}\right)\right) \nonumber\\
&=\nabla \bar{\mathit{g}}\left(\boldsymbol{\phi}_{k}\right)-\Re\left\{\nabla \bar{\mathit{g}}\left(\boldsymbol{\phi}_{k}\right) \odot \boldsymbol{\phi}_{k}^{*}\right\} \odot \boldsymbol{\phi}_{k},
\end{align}
where $\operatorname{Proj}\left ( \nabla \bar{\mathit{g}}\left(\boldsymbol{\phi}_{k}\right)\right )$ denotes the orthogonal projection of the Euclidean gradient $\nabla \bar{\mathit{g}}\left(\boldsymbol{\phi}_{k}\right)$ onto $\mathcal{T}  _{\boldsymbol{\phi}_{k}} \mathcal{M}_{cc}^{N}$. Through numerical computation, we can obtain the Euclidean gradient $\nabla \bar{\mathit{g}}\left(\boldsymbol{\phi}_{k}\right)$ at point $\boldsymbol{\phi}_{k}$ as follows:
\begin{align}\label{51}
 \nabla \bar{\mathit{g}}\left(\boldsymbol{\phi}_{k}\right)=2\mathbf{A}\boldsymbol{\boldsymbol{\phi}}_{k}+\mathbf{c}+\mathbf{w}^{\mathrm{H}}.
\end{align}
Since the tangent vector obtained from the Riemannian gradient may not lie on the surface of the manifold, we need a retraction operation to map this vector to a new point on the manifold. This map procedure can be expressed as follows: 
\begin{align}\label{52}
	\boldsymbol{\phi}_{k+1} & =\operatorname{Ret}_{\boldsymbol{\phi}_{k}}\left(\delta_{k} \boldsymbol{\gamma}_{k}\right) \nonumber\\
	& \triangleq \mathcal{T}_{\boldsymbol{\phi}_{k}} \mathcal{M}_{c c}^{N} \mapsto \mathcal{M}_{c c}^{N}: \delta_{k} \boldsymbol{\gamma}_{k} \mapsto \operatorname{unit}\left(\frac{\boldsymbol{\phi}_{k}+\delta_{k} \boldsymbol{\gamma}_{k}}{\left|\boldsymbol{\phi}_{k}+\delta_{k} \boldsymbol{\gamma}_{k}\right|}\right),
\end{align}
where $\boldsymbol{\gamma}_{k}$ denotes the search direction passing through $\boldsymbol{\phi}_{k}$, $\delta_{k}$ is the Armijo backtracking line search step size, $\mapsto$ denotes the operation of mapping, and $\operatorname{unit}$ denotes the normalization of vector.

With conjugate gradient descent method, the $(k+1)$th search direction in Euclidean space is
\begin{align}\label{53}
\boldsymbol{\gamma}_{k+1}=-\nabla \bar{\mathit{g}}\left(\boldsymbol{\phi}_{k+1}\right)+\mathfrak{d}_{k} \boldsymbol{\gamma}_{k},
\end{align}
where $\mathfrak{d}_{k}$ is the Polak-Ribiere parameter of the $k$th iteration \cite{absil2008optimization}. Since $\boldsymbol{\gamma}_{k+1}$ and $\boldsymbol{\gamma}_{k}$ lie on different tangent spaces and cannot be conducted directly  with mathematical computation, $\boldsymbol{\gamma}_{k}$ at point $\boldsymbol{\phi}_{k}$ needs to be transported to $\boldsymbol{\gamma}_{k+1}$ at point $\boldsymbol{\phi}_{k+1}$, the transport operation is given by
\begin{align}\label{54}
\mathcal{T}_{\boldsymbol{\phi}_{k}\mapsto\boldsymbol{\phi}_{k+1}}\left(\boldsymbol{\gamma}_{k}\right)&=\mathcal{T}_{\boldsymbol{\phi}_{k}} \mathcal{M}_{c c}^{N} \mapsto \mathcal{T}_{\boldsymbol{\phi}_{k+1}}\mathcal{M}_{c c}^{N}\nonumber\\
&=\boldsymbol{\gamma}_{k}\mapsto\boldsymbol{\gamma}_{k}-\Re\left\{\boldsymbol{\gamma}_{k} \odot \boldsymbol{\phi}_{k}^{*}\right\} \odot \boldsymbol{\phi}_{k}.
\end{align}
With the Riemannian gradient, the $(k+1)$th
search direction in the manifold space can be updated by
\begin{align}\label{55}
\boldsymbol{\gamma}_{k+1}=-\nabla _{\mathcal{M}} \bar{\mathit{g}}\left(\boldsymbol{\phi}_{k+1}\right)+\mathfrak{d}_{k} \mathcal{T}_{\boldsymbol{\phi}_{k}\mapsto\boldsymbol{\phi}_{k+1}}\left(\boldsymbol{\gamma}_{k}\right).
\end{align}
\begin{algorithm}[t] 
	\caption{RCG algorithm for solving (P6$''$)}
	\label{RCG}
	\begin{algorithmic}[1]
		\STATE Initialize with $k=0$, $\boldsymbol{\phi}_{0} \in \mathcal{M}_{c c}^{N}$, and $\boldsymbol{\gamma}_{0}=-\nabla _{\mathcal{M}} \bar{\mathit{g}}\left(\boldsymbol{\phi}_{0}\right)$.  
		\STATE \textbf{repeat}
		\STATE Choose the step size $\delta_{k}$ by backtracking line search;
		\STATE Update point $\boldsymbol{\phi}_{k+1}$ by \eqref{52};
		\STATE Update the Riemannian gradient $\nabla _{\mathcal{M}} \bar{\mathit{g}}\left(\boldsymbol{\phi}_{k+1}\right)$ by \eqref{50} and \eqref{51};
		\STATE Update the transport operation $\mathcal{T}_{\boldsymbol{\phi}_{k}\mapsto\boldsymbol{\phi}_{k+1}}\left(\boldsymbol{\gamma}_{k}\right)$ by \eqref{54};
		\STATE Choose the Polak-Ribiere parameter $\mathfrak{d}_{k}$;
		\STATE Update the new search direction $\boldsymbol{\gamma}_{k+1}$ by \eqref{55};
		\STATE $k=k+1$;
		\STATE \textbf{until} $\left\|\nabla _{\mathcal{M}} \bar{\mathit{g}}\left(\boldsymbol{\phi}_{k+1}\right)\right\|^{2} \leq \epsilon $.
		\STATE  \textbf{Return} $\ddot{\boldsymbol{\phi}}^{\mathcal{F}_{2}}=\boldsymbol{\phi}_{k+1}$ 
	\end{algorithmic}
\end{algorithm}

The specific procedure of RCG algorithm to solve (P6$''$) is summarized in Algorithm 1, which continuously updates the gradient search direction based on \eqref{55} until the Riemannian gradient $\nabla _{\mathcal{M}} \bar{\mathit{g}}\left(\boldsymbol{\phi}_{k+1}\right)$ converges.  The complexity of RCG algorithm is mainly dominated by calculating
the Euclidean gradient, which is $O(N^2)$. Thus, the computational complexity of RCG algorithm is $O(I_RN^2)$ where $I_R$ represents 
the iteration times required by the RCG algorithm. According to~\cite{boyd2004convex}, the
complexity of SDR method is $O(N^{3.5})$. It can
be observed that the complexity of RCG algorithm is lower than that of the SDR method.

\par 3) If $\mathrm{\phi}_{n} \in \mathcal{F}_{3}$, the constraint of $\text { (P1) }$ is  
\begin{align}
\mathrm{\phi}_{n}=e^{j \mathrm{\theta}_{n}},\theta_{n} \in\left\{0, \frac{2 \pi}{\tau}, \cdots, \frac{2 \pi(\tau-1)}{\tau}\right\}.
\end{align}
We adopt NPP method to deal with the discrete phase and choose to obtain the discrete phase RC from the projection of the continuous phase RC due to the relatively low computational complexity \cite{guo2019weighted}. We denote by $\ddot{\Omega}^{\mathcal{F}_{2}}_{n}$ the angle of the
optimal solution to (P6$''$). Then, for the case where $\phi_{n} \in \mathcal{F}_{3}$, we can obtain the subptimal solution to the discrete phase as follows:  
\begin{align}\label{57}
\ddot{\phi}^{\mathcal{F}_{3}}_{n}=e^{j \ddot{\Omega}^{\mathcal{F}_{3}}_{n}},\quad \forall \mathit n\in \mathcal{N},
\end{align}
where
\begin{align}
\ddot{\Omega}^{\mathcal{F}_{3}}_{n}=\arg \min _{\theta_{n} \in\left\{0, \frac{2 \pi}{\tau}, \cdots, \frac{2 \pi(\tau-1)}{\tau}\right\}}\left|\theta_{n}-\ddot{\Omega}^{\mathcal{F}_{2}}_{n}\right| ,\quad \forall \mathit n\in \mathcal{N}.
\end{align}
\subsection{Overall Algorithm}
Based on previous discussions, the AO algorithm to solve (P1) is shown in Algorithm 2. The specific steps of the algorithm are as follows: Firstly, the algorithm starts with initial feasible points $\mathbf p_{1}^{(0)}$ and $\boldsymbol{\phi}^{(0)}$. In the next step, with a fixed solution $\mathbf p_{1}^{(i-1)}$ and $\boldsymbol{\phi}^{(i-1)}$ from the $(i-1)$th iteration, the update of $\mathbf p_{1}^{(i)}$ and $\boldsymbol{\phi}^{(i)}$ is performed alternately using fractional programming techniques in the $i$th iteration. Note that the objective function of (P1) is monotonically non-decreasing, hence Algorithm 2 will eventually converge\cite{jiang2021joint}.

\begin{algorithm}[t] 
	\caption{Alternating Optimization for Solving (P1)}
	\label{AO}
	\begin{algorithmic}[1]
		\STATE Initialize with feasible points $\mathbf p_{1}^{(0)}$ and $\boldsymbol{\phi}^{(0)}$.  
		\STATE \textbf{repeat}
		\STATE Update $\mathbf q^{(i)}$ by \eqref{25};
		\STATE Update power allocation vector $\mathbf p_{1}^{(i)}$ by \eqref{30};
		\STATE Update $\mathbf l^{(i)}$ by \eqref{35};
		\STATE Update RC vector $\boldsymbol{\phi}^{(i)}$ by solving problem (P1);
		\STATE $i=i+1$;
		\STATE \textbf{until} the value of the objective function of (P1) reaches convergence.
		\STATE  Return the optimal solution $\ddot{\mathbf p}_{1}=\mathbf p_{1}^{(i)}$ and $\ddot{\boldsymbol{\phi}}=\boldsymbol{\phi}^{(i)}$.
	\end{algorithmic}
\end{algorithm}

\section{Simulation}\label{sec:Simulation}
In this section, we give simulation results under different conditions to validate the SIC capability and the capacity gain of RIS-assisted FD-OFDM communications over HD-OFDM communications.
\par We assume that the transmit and receive antenna of each FD device is located at $(-0.02,0,0.04)$~m and $(0.02,0,0.04)$~m, respectively. The number of OFDM subcarriers, the maximum delay spread, and the length of CP is set to $M = 128$, $L_{\rm max}=5$, and $M_{\rm cp}=5$, respectively. %We consider an OFDM system with $M = 128$ subcarrriers, the maximum delay spread $L_{max}=5$, and the length of CP $M_{cp}=5$.
We set the distance between two FD devices to 1000 m, the area of each unit cell under the $m$th subcarrrier to $R_m=(\lambda_m / 5)^{2},\forall \mathit m\in \mathcal{M} $, and the reflection efficiency of each RIS to $\eta=\beta=0.8$. %$\tau=4$, 
Also, the total power of each transmitter is set to $P_1=P_2=P$, the noise power is set to $\sigma_{1}^{2}=-110$ dBm, %the number of Gaussian randomization is set to $I=100$, 
and $\epsilon=10^{-7}$. The pathloss %at distance b 
is given by $L=\zeta s^{-\partial  }$, where $s$ is the distance and $\zeta$ is the pathloss at the reference distance of 1~m, which is set to -30~dB with the pathloss exponent of $\partial=2$. If $\mathbf{\Gamma } \in \{\mathit{h}_{m,\rm T1},\mathit{h}_{m,\rm T2}\}, \forall \mathit m\in \mathcal{M} $, we set $K_{\rm r}=6$. If $\mathbf{\Gamma } \in
  \{\mathbf{g}_{m,12},\mathbf{g}_{m,21},\mathbf{h}_{m,12},\mathbf{h}_{m,21}\}, \forall \mathit m\in \mathcal{M} $, we set $K_{\rm r}=9$. The channels between elements of a RIS and an antenna are assumed to be independent and identically distributed. However, the channels from a RIS to different antennas are assumed to be non-identical. 
  Apart from our schemes developed in Section III, we also introduce benchmark scheme for comparison. The descriptions of our developed and the benchmark scheme are as follows:
\begin{itemize}
  	\item[1)]
  	\textbf{Ideal case:} in this scheme, both the RC matrix of the RIS and power allocation vector are optimized to maximize the SIC capability under the case of $\mathrm {\phi}_{n} \in \mathcal{F}_{1}$.
\end{itemize}
\begin{itemize}
	\item[2)]
	\textbf{Continuous phases:} we apply RCG algorithm to optimize the RC matrix where $\mathrm {\phi}_{n} \in \mathcal{F}_{2}$.
\end{itemize}
\begin{itemize}
	\item[3)]
	\textbf{Discrete phases:} when the phase shift $\mathrm {\phi}_{n}$ belongs to the set $\mathcal{F}_{3}$, we set the number of discrete phases to be $\tau = 2, 4,$ and $8$ for comparison.
\end{itemize}
\begin{itemize}
	\item[4)]
	\textbf{Random phases:} 
	we set the random phases scheme as the benchmark scheme. In our assumption, the RC has a random phase that is randomly selected from [0, $2\pi$], while the amplitude is fixed to one~\cite{zhang2021joint}.
\end{itemize}
  
  %We compare the SIC performance of the proposed method with that of antenna cancellation  scheme \cite{choi2010achieving}, which employed asymmetric placement of TX antennas to generate a $\rm \pi$ phase shift between the receive signals at the RX.
  %uses signal reversal based on  Balun transceiver to suppress SI.
  
\begin{figure}[t]
  	\centering
  	\vspace{-10pt}
  	\includegraphics[scale=0.64]{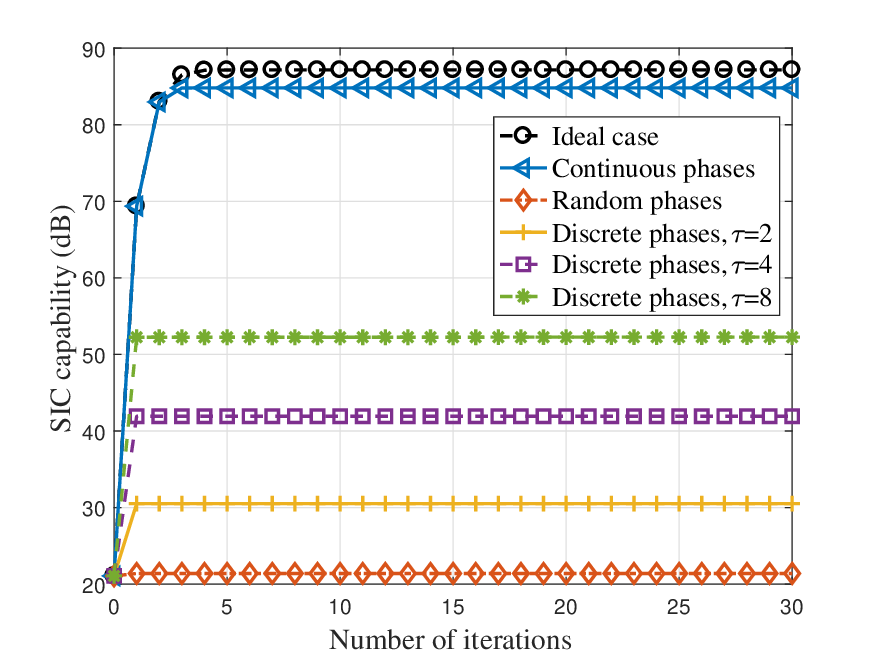}
  	\caption{The convergence behavior of Algorithm 2.} \label{fig:convergence}
  	\vspace{-10pt}
\end{figure} 
\begin{figure}[htbp]
	\centering
	\vspace{-10pt}
	\subfigbottomskip=5pt %??????????????????????????????
	\subfigcapskip=5pt %?????????????
	\subfigure[SIC capability. ]{
		\begin{minipage}{1\linewidth}\label{fig:SIC_N}
			\centering
			\includegraphics[scale=0.64]{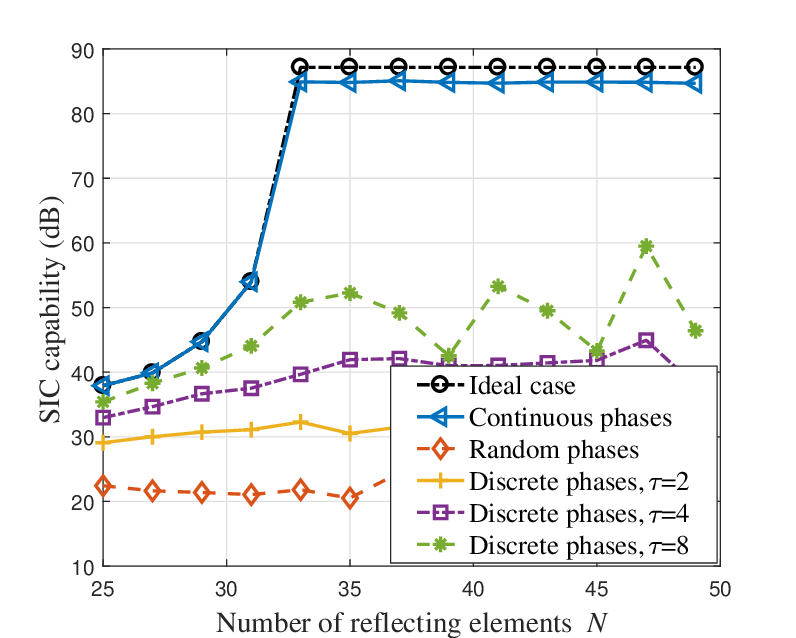}
		\end{minipage}
	}
	\\
	\vspace{-10pt}
	\subfigure[Capacity gain.]{
		\begin{minipage}{1\linewidth}\label{fig:capacity_N_compare}
			\centering
			\includegraphics[scale=0.64]{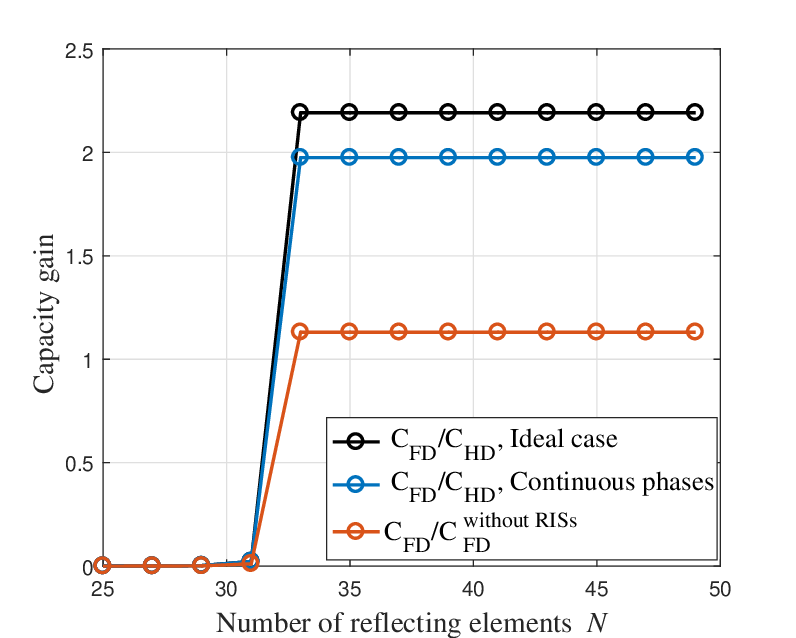}
		\end{minipage}
	}
	\centering
	\caption{The SIC capability and capacity gain with different values of $N$.} \label{fig:SIC_N&cap_N}
	\vspace{-10pt}
\end{figure}
\par First, we evaluate the convergence behavior of Algorithm~2. The number of RIS elements is set as $N=35$, the transmit power is set as $P=0$~dBm, and the signal bandwidth is set as $B=20$~MHz. If not mentioned otherwise, these parameter settings remain constant throughout the subsequent discussion. We initialize the power allocation vector $\mathbf p_{1}^{(0)}$ as a column vector of zeros and the RC vector $\boldsymbol{\phi}^{(0)}$ as a column vector of ones, except for the random phases scheme where the phase of each unit cell is initialized as a random value selected from [0, $2\pi$]. Fig.~\ref{fig:convergence} shows the SIC capability versus the number of iterations in different conditions. We can observe from Fig.~\ref{fig:convergence} that all curves converge  within 5 iterations. To be specific, the ideal case and the continuous phases case converge within 5 iterations, while the other cases converge within 1 iteration.  For the ideal phases case, the power allocation vector and RC vector are obtained by solving the KKT conditions, which can be expressed in closed-form expression. Thus, the optimal solution can be obtained in each iteration, resulting in convergence. In addition, the RC vector of the continuous phases case is optimized by the RCG algorithm, which also offers a fast convergence. Thus, the continuous phases curve can converge within 5 iterations. The discrete phases case converges faster than the ideal and continuous phases cases for two reasons. First, the feasible domain of the discrete phases case is small, we only need to choose one of several discrete values as the phase shift. Second, the discrete phases are directly projected from the continuous phases. Thus, we only need to optimize the power allocation without optimizing the RC. %This results in faster convergence compared to the continuous and ideal phases.
Similarly, the random phases are generated randomly, and only the power allocation is optimized, resulting in the same fast convergence speed as the discrete phases. Since both the amplitude and phase of each RIS unit cell can be independently and continuously controlled, the ideal case can achieve the optimal performance, which is consistent with the discussion in Section II.%Furthermore, it can be observed from the figure that although the SDR method converges faster, its algorithm performance is inferior to that of the NPP method, which is consistent with the discussion in Section III.

%Figure~\ref{fig:SIC_N&cap_N} shows the SIC capability and the capacity gain which denote by $C_{\rm FD}/C_{\rm HD}$ with different values of $N$.
In Fig.~\ref{fig:SIC_N&cap_N}, the SIC capability and capacity gain are depicted with different values of $N$. As shown in Fig.~\ref{fig:SIC_N}, the SIC capabilities of the ideal case and the continuous phases case increase monotonically as the value of $N$ increases and it can be larger than 85~dB. When $N$ is larger than 32, both the ideal case and the continuous phases case exhibit a consistent SIC capability that remains stable as $N$ increases. The curve of the ideal case can remain stable because the modules of the RC are not fixed to one. This stability of the continuous phases case is achieved because the optimized power allocation vector compensates for the unit modulus constraint and ensures that the SIC capability does not decrease as $N$ increases. %The SIC capability of the discrete phase improves as the value of $\tau$ increases, since larger $\tau$ values result in candidate phases that are closer to ideal phases. The SIC capability with discrete phases is weaker than that with "Ideal case" and "Continuous phases, NPP" due to the phase differences between candidate phases and ideal phases. The SIC capability reduction resulting from the phase differences can be compensated by increasing the number of reflecting elements. 
Increasing the value of $\tau$ leads to an enhancement in the SIC capability of the discrete phases because the candidate phases approach the ideal phases more closely as $\tau$ increases. However, as compared with the cases of ideal phases and continuous phases, the SIC capability of the discrete phases is weaker due to the phase differences between candidate phases and ideal phases. In the case of discrete phases, the SIC capability fluctuates with the increase of $N$ due to the varying phase differences. At times, these phase differences increase, leading to a decrease in SIC capability. Conversely, there are instances where the phase differences decrease, resulting in an increase in SIC capability. Thus, the fluctuation of SIC capability with $N$ can be attributed to the changing phase differences. This reduction in SIC capability resulting from these phase differences can be compensated by increasing the number of reflecting elements.
The SIC achieved with random phases is approximately 20~dB,  which is significantly lower than that of the ideal case and the continuous phases case. This result implies the importance of phase shift design in SIC. 

The comparison of our proposed RIS-assisted FD-OFDM system with the FD-OFDM system without RISs is shown in Fig.~\ref{fig:capacity_N_compare}. The system capacity of FD-OFDM system without RISs are obtained by replacing~\eqref{Capacity_FD} with
\begin{align} \label{Capacity_FD_withoutRIS}
C_{\rm FD}^{\rm N-RISs}=\frac{1}{M+M_{cp}} \sum_{m=0}^{M-1} \mathrm{\log} _{2}\hspace{-1mm}\left(1+\frac{\left|\mathit{h}_{m,\rm T2}\right|^{2} p_{2,m}}{\kappa\left|\mathit{h}_{m,1}\right|^{2} p_{1,m}+\sigma_{1}^{2}}\right),
\end{align}
where $0 \le \kappa \le 1$ is the SIC coefficient \cite{cheng2013optimal, yang2017performance, mahady2022energy}. When $\kappa = 0$, the SI is perfectly canceled. On the other hand, when $\kappa = 1$, SIC techniques are invalid. To better clarify the capacity gain of the proposed
RIS-assisted FD-OFDM communication system over FD-OFDM system without RISs, we set $\kappa = 1/(87 \rm ~dB)$ where 87 dB is the SIC capability of RIS-assisted FD-OFDM system. 
As illustrated in Fig.~\ref{fig:capacity_N_compare}, the capacity gain is initially close to zero due to the strong SI. However, the capacity gain gradually increases as $N$ increases, and then ultimately reaches a steady state. It is worth noting that although RISs are deployed to cancel the SI, the RIS-assisted link indirectly strengthens the  desired signal. Thus, the capacity gain of the proposed RIS-assisted FD-OFDM communication system over HD-OFDM system without RISs can be more than 2.25 times. With the same SIC capability, we can observe that the capacity gain of RIS-assisted FD-OFDM communication system over FD-OFDM system without RISs is 1.13, which confirms that RIS can significantly improve the capacity as compared with FD-OFDM communication system without RISs. This is because the RIS-assisted link not only cancels the SI but also strengthens the desired signal. The results in Fig.~\ref{fig:SIC_N&cap_N} demonstrate the feasibility of deploying RIS for SIC and provide a novel approach for SIC in the propagation domain for FD communications. 
% the SIC capability of antenna cancellation  scheme is 30~dB, while %compared with the Balun scheme which signal inversion uses a simple design
%based on a balanced/unbalanced (Balun) transformer.
 %the SIC capability of RISs-assisted scheme increases monotonically as the values of $P$ and $N$ increase
 %is continuously improved with the increase of $N$  and $P$ 
 %and it can be larger than 80~dB. If $N\ge36$,  the SIC capability is only affected by the transmit power due to the impact of noise.
  %Also, as shown in Fig.~\ref{fig:capacity_N}, the system capacity before SIC is close to zero because of the strong SI. With the proposed RISs based SIC scheme, the capacity increases as the value of $N$ increases and converges if $N>35$. %then the capacity double the capacity of HD.
%It is worth noting that although RISs are deployed to cancel the SI, the RIS-assisted link indirectly strengthens the  desired signal. Thus, the capacity gain of the proposed RISs-assisted FD communication system over HD system without RISs can be more than 2.5 times. %the capacity of the proposed RISs-assisted FD communication system can be increased by about 2.5 times as compared with HD system. %the relatively small gain to the desired signal results in no further improve in the capacity.
%the gain of the desired signal is relatively small and there is no further improvement in capacity.
%the relatively small gain to the desired signal results in no further improvement in capacity.
%the relatively gain to the desired signal results in  further improvement in capacity.
 \begin{figure}[t]
 	\centering
 	\vspace{-10pt}
 	\setlength{\abovecaptionskip}{0pt} %?????????
 	\setlength{\abovecaptionskip}{2pt} %?????????
 	\includegraphics[scale=0.64]{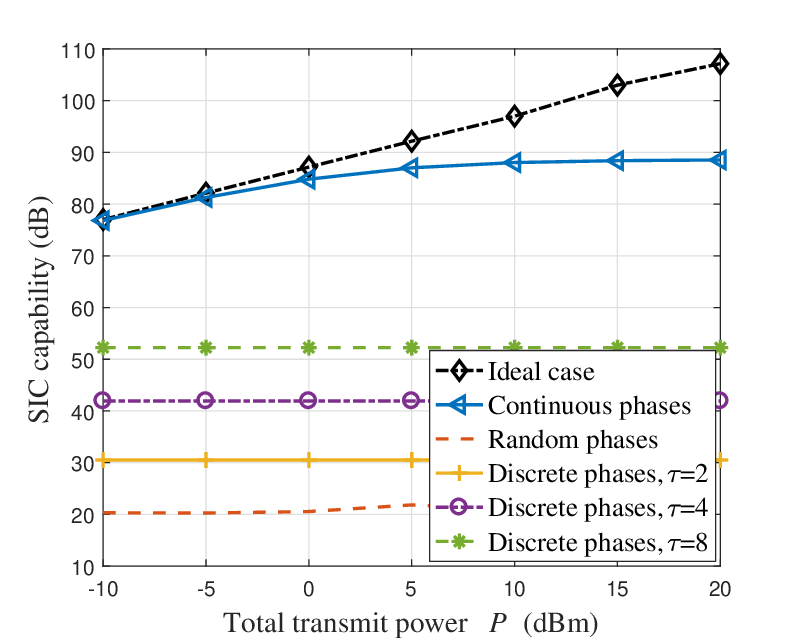}
 	\caption{The SIC capability with different values of $P$.} \label{fig:SIC_P}
 	\vspace{-10pt}
 \end{figure}
\begin{figure}[t]
	\centering
	\vspace{-10pt}
	\setlength{\abovecaptionskip}{0pt} %?????????
	\setlength{\abovecaptionskip}{2pt} %?????????
	\includegraphics[scale=0.64]{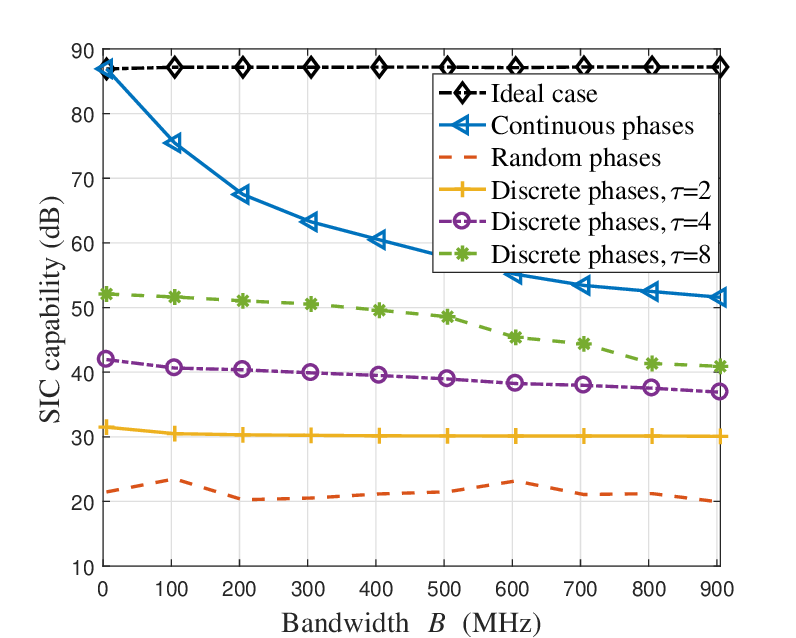}
	\caption{The SIC capability with different values of $B$.} \label{fig:SIC_B}
	\vspace{-10pt}
\end{figure}

Figure~\ref{fig:SIC_P} plots the SIC capability corresponding to the total transmit power $P$. Fig.~\ref{fig:SIC_P} illustrates that as $P$ increases, the SIC capabilities of the ideal case and the continuous phases case increase accordingly. However, the SIC capability of the continuous phases case is weaker than that of the ideal case due to the unit modulus constraint imposed on each RIS units, which limits its performance. In cases of discrete phases and random phases, the SIC capabilities are not affected by transmit power. Hence, increasing $P$ fails to improve the SIC capability. This outcome is attributed to the inadequate cancellation of SI. The residual SI remains very strong, which diminishes the significance of transmit power in increasing the SIC capability.

In Fig.~\ref{fig:SIC_B}, the SIC capability of the proposed scheme with different values of signal bandwidth $B$ is presented. Fig.~\ref{fig:SIC_B} shows that the SIC capabilities of all curves, except for the ideal case and the random phases case, decreases as $B$ increases. The reason for the curve of the ideal case remains unchanged is that the SI is effectively canceled, so it is not affected by bandwidth. The reason why the case of random phases is not affected by bandwidth, as same as Fig.~\ref{fig:SIC_P}, is the strong residual SI. The performance of the continuous phases case is similar to the ideal case when $B=5$~MHz, and then the SIC capability drops to $80$ dB when $B=55$~MHz due to the unit-modulus constraint. %RIS can only bring the same phase shift on different subcarriers%
In practical design of RIS phase shift, passive beamforming with phase gradient metasurfaces can be used to cancel SI. The phase difference between different frequency points in the feasible signal bandwidth range is relatively constant, which implies that signal bandwidth has no impact on the SIC capability. 
The aforementioned development is encouraging, as it suggests that the implementation of RISs to cancel SI in FD broadband communications is feasible.

We can utilize phase discretization to simulate the phase errors of continuous phases. As the number of discrete phases increases, the phase errors decrease. When the number of phases approaches infinity, the phase errors become zero. %This inaccuracy results in phase errors, which can cause an increase in sidelobe levels and a decrease in performance gain~\cite{yang2017study}. 
The phase errors are defined by
\begin{align}\label{eq:phase_errors}
\Delta \Omega_{n}=\left|\ddot{\Omega}^{\mathcal{F}_{3}}_{n}-\ddot{\Omega}^{\mathcal{F}_{2}}_{n}\right| \approx \frac{360^{\circ}}{\tau }  ,\quad \forall \mathit n\in \mathcal{N},
\end{align}
where $\ddot{\Omega}^{\mathcal{F}_{2}}_{n}$ denotes the ideal phase of the continuous phases case and $\ddot{\Omega}^{\mathcal{F}_{3}}_{n}$ is the actual phase of the $n$th unit cell selected from the $\tau$ available phases.

\begin{figure}[t]
	\centering
	\vspace{-10pt}
	\setlength{\abovecaptionskip}{0pt} %?????????
	\setlength{\abovecaptionskip}{2pt} %?????????
	\includegraphics[scale=0.64]{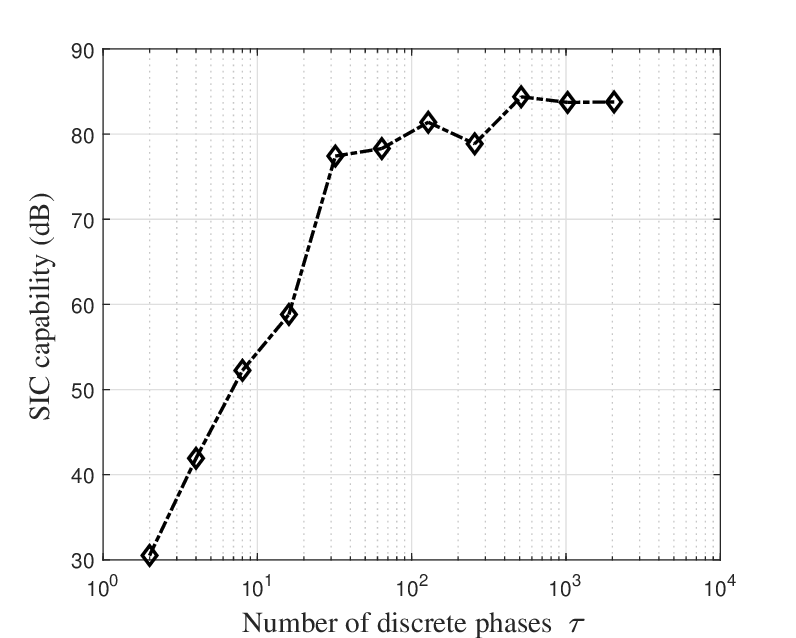}
	\caption{The SIC capability with with different values of $\tau$.} \label{fig:phase_errors}
	\vspace{-10pt}
\end{figure}

Figure~\ref{fig:phase_errors} presents the SIC capability corresponding to different numbers of discrete phases $\tau$. We set $\tau$ to increase by $2^\omega $, where $\omega$ is the resolution bits for the discrete phases. 
As shown in Fig.~\ref{fig:phase_errors}, the SIC capability gradually increases as $\tau$ increases. When $\tau\ge512$, the SIC capability of the discrete phases is close to that of the continuous phases, with phase errors of about $0.7^{\circ}$. When  $\tau\ge128$, i.e., the phase error is $2.8^{\circ}$, which is larger than the actual engineering design error~\cite{ali2020integration, wang2019phase}. However, the SIC capability only decreases by 5 dB, and there is still 80 dB of SIC capability, which is enough to ensure the performance of the RIS.
%The SIC capability can reach 80~dB, which is only 5~dB lower than that of the continuous phases.
 Thus, the phase accuracy of the proposed RIS-assisted FD-OFDM communication system is sufficient. 

The size of RIS is the primary limiting factor for deploying it inside a device. For the considered 5.8 GHz band, RIS can be deployed inside a device since its size can be relatively small. %The size of RIS has emerged as the primary limiting factor for deploying them within FD device, owing to the relatively small size of such devices. 
	The RIS unit side length can typically be selected between $\lambda/10$ and $\lambda/2$~\cite{tang2021wireless, bjornson2020power}. We can set the RIS unit length to $\lambda/5$. For the case where the operating frequency is $f=5.8$ GHz, the wavelength is $\lambda=c/f=0.0517$~m, and the area of RIS unit cell is $\left ( \lambda/5 \right )^2\approx \left ( 0.01 \right )^2 \rm m^2$. As shown in Fig. 5(a) of this paper, a total of 36 RIS units are required to achieve maximum SIC capability. Therefore, the size of RIS is approximately $60 \times  60~\rm mm^2$, which can be deployed inside some terminal products, such as HUAWEI CPE series.

\section{Conclusions}\label{sec:Conclusion}
In this paper, we proposed and analyzed a novel SIC scheme based on RIS for FD-OFDM communications. 
%We first analyzed the near-field behavior of the RIS and then an optimization problem was formulated to maximize the SIC capability. In the final stage of our study,  we utilized the AO algorithm to jointly optimize the power allocation vector and RC matrix of the RIS in three cases, with the goal of minimizing residual SI.
First, we analyzed the near-field behavior of the RIS and found the relationships between near-field channel and performance optimization. We then formulated an optimization problem to maximize the SIC capability by controlling the RC of the RIS and allocating the transmit power of the device. Finally, we utilized the AO algorithm to solve the problem in ideal case, continuous phases, and discrete phases. %Through this approach, we were able to identify the most efficient and effective use of RIS in the context of SIC.
Simulation results showed that the proposed RIS-based SIC scheme can cancel the SI by more than 85~dB in the ideal case and the continuous phases case. Furthermore, we evaluated the impact of phase errors caused by small-sized RIS in SIC capability. The unsignificant difference in SIC capability between two cases indicates that the low-complexity RIS with fixed amplitude of 1 is a viable option. %More importantly, there is not a significant difference in SIC capability between ideal case and continuous phases. This indicates that, in practical RIS implementation, using a realistically designed RIS with a fixed amplitude of 1 instead of an ideal RIS with independent and continuous control over each RIS unit cell's amplitude and phase is a viable option.
Also, the capacity gain of the proposed RIS-assisted FD-OFDM system over HD-OFDM system without RISs can be more than 2.25 times due to the enhancement for the effective path of the desired signal. %We have demonstrated that in the optimization problem of maximizing SIC capability, the performance of the NPP method outperforms that of the SDR method. 
These encouraging results suggest that RIS can provide an attractive way to achieve FD communications.%the system capacity can be increased by about 2.5 times as compared with HD system due to the enhancement of effective path of desired signal. These encouraging

%can significantly mitigate SI and effectively enhance system capacity. 

%\begin{appendices}
%\section{Proof for Upper and Lower Limits of $\widehat{C}_{\rm{OAM}}$}\label{pro:COAM2_upper_lower}
%
%\end{appendices}

\bibliographystyle{IEEEtran}
\bibliography{References}

\end{document}